\documentclass[12pt,a4paper]{article} 
\usepackage[utf8]{inputenc} 
\usepackage[dvipsnames]{xcolor}
\usepackage{pdfpages}

\usepackage[a4paper, total={6.2in, 9.2in}]{geometry}

\usepackage{mathrsfs}
\usepackage{graphicx}
\usepackage{subcaption} 

\usepackage{hyperref}
\usepackage{latexsym}

\usepackage{bm} 
\usepackage[dvipsnames]{xcolor}

\usepackage{authblk}

\newcommand{\yb}[1]{\boldsymbol{#1}}
\newcommand{\yfrac}[2]{\displaystyle{\frac{#1}{#2}}}
\newcommand{\yd}[2]{\yfrac{ d #1}{d #2}}
\newcommand{\vecb}[1]{{\boldsymbol #1}}
\def \d {\mathrm{d}}
\newcommand{\fpar}[2]{\frac{\partial{#1}}{\partial{#2}}}

\newcommand{\aeq}{\begin{equation}}
\newcommand{\eeq}{\end{equation}}
\newcommand{\aeqn}{\begin{eqnarray}}
\newcommand{\eeqn}{\end{eqnarray}}
\newcommand{\aeqns}{\begin{eqnarray*}}
\newcommand{\eeqns}{\end{eqnarray*}}

\newcommand{\yB}{\yb{B}}

\begin{document}
\title{\vspace{-1.0cm} \large \bf Linear gyrokinetic investigation of the geodesic acoustic modes in realistic tokamak configurations}
\author[1]{\normalsize I. Novikau}
\author[1]{A. Biancalani}
\author[1]{A. Bottino}
\author[1]{G. D. Conway}
\author[2]{\"O. D. G\"urcan}
\author[1]{P. Manz}
\author[2]{P. Morel}
\author[1]{E. Poli}
\author[1]{A. Di Siena}
\author[1]{the ASDEX Upgrade Team}
\affil[1]{Max-Planck-Institut f\"ur Plasmaphysik, 85748 Garching Germany}
\affil[2]{LPP, CNRS, {\'E}cole polytechnique, UPMC Univ. Paris 06, Univ. Paris-Sud, Observatoire de Paris, Universit{\'e} Paris-Saclay, Sorbonne Universit{\'e}s, PSL Research University, F-91128 Palaiseau, France}
\affil[ ]{\textit{ivannovi@ipp.mpg.de}}
\date{}
\maketitle
\begin{abstract}
Geodesic acoustic modes (GAMs) are studied by means of the gyrokinetic global particle-in-cell code ORB5.
Linear electromagnetic simulations in the low-$\beta_e$ limit have been performed, in order to separate acoustic and Alfv\'enic time scales and obtain more accurate measurements.
The dependence of the frequency and damping rate on several parameters such as the safety factor, the GAM radial wavenumber and the plasma elongation is studied. 
All simulations have been performed with kinetic electrons with realistic electron/ion mass ratio. 
Interpolating formulae for the GAM frequency and damping rate, based on the results of the gyrokinetic  simulations, have been derived. 
Using these expressions, the influence of the temperature gradient on the damping rate is also investigated. Finally, the results are applied to the study of a real discharge of the ASDEX Upgrade tokamak.
\end{abstract}

\section{Introduction} 
The ion heat transport  in the plasma core is governed by turbulence formed by a class of microinstabilities such as toroidal ion temperature gradient (ITG) driven modes \cite{Horton99}. 
ITG turbulence is known to self-organize to form macroscopic structures \cite{Manz09}. 
These structures take the form of a macroscopic radial electric field which depends only on the radial coordinate. 
$E \times B$ poloidal flows associated with this electric field are referred to as zonal flows (ZFs) \cite{Hasegawa79,Rosenbluth98,Diamond05,Gurcan15}. 

The action of the toroidal magnetic field curvature on the ZF gives rise to oscillations of the radial electric field. 
These oscillations of the ZFs are called geodesic acoustic modes (GAMs) \cite{Winsor68,Zonca08}.  
The modes are observed predominantly in the edge region of the tokamak plasmas with characteristic frequency of the order of the sound frequency $\sim c_s/R$, where $c_s = \sqrt{T_e/m_i}$ is the sound speed, $R$ is the major radius. One of the main linear damping mechanisms for the stationary ZF are collisional processes and for the GAM it is a collisionless wave-particle interaction, namely the Landau damping, and collisional damping at the very edge of the plasma, where equilibrium temperatures drastically decrease \cite{Gao2013}. A recent comparison of collisionless and collisional damping of GAMs, 
using existing analytical theories, for experimentally relevant plasmas 
was done in Ref. \cite{Simon2016}.

The importance of the ZF is that they can regulate the drift-wave (DW) turbulence \cite{Biglari90}. 
But it is still a question how the GAMs influence the ZF efficiency of the DW suppression \cite{Zonca08, Miki11, Scott03}. 
On the other hand, the development of zonal structures can play a key role in the transition from the low to the high confinement regime (L-H transition) \cite{Manz12}. 
In Ref. \cite{Conway11} interaction of the mean and oscillatory poloidal flows with the turbulence were experimentally observed. The turbulence suppression by the ZFs was observed in experiments described in Ref. \cite{Schmitz12}. On the other hand, in Ref. \cite{Cavedon16} the role of the mean flow in the dynamic evolution towards the H-mode is emphasized. 
In Ref. \cite{Miki2011} two predators - one prey system, including ZF, GAM and turbulence, was developed to study transitions between states with different combinations of the ZF and GAM.

In this paper, we investigate the GAM frequency and collisionless damping rate, carrying out linear collisionless simulations with kinetic electrons.
The electromagnetic global gyrokinetic particle-in-cell code ORB5 is used \cite{Jolliet07,Bottino15}.
As it has been reported previously \cite{Zhang10,Biancalani17}, models, numerical or analytical, derived with adiabatic electrons, result in considerably smaller GAM damping rate in comparison to simulations performed with kinetic electrons.
By adiabatic electron models, we mean here models treating the $m\ne 0$ component of the electrons as adiabatic, and setting the zonal component of the electron density perturbation to zero.
In simulations considered in this paper, electrondocumentclasss are treated drift-kinetically, and a realistic ion-electron mass ratio is used.
Moreover, to study the influence of the plasma elongation on the GAM dynamics, magnetic equilibria with realistic plasma shapes are considered. 
To summarize the results obtained in different plasma regimes, interpolating formulae for the GAM frequency and damping rate, based on the gyrokinetic simulations with ORB5, are derived.

Due to the so-called phase mixing effect, the GAM damping rate is increased in the presence of a temperature gradient or the safety factor profile \cite{Zonca08,Palermo16,Biancalani16}.
This effect arises when the damping rate of the wave depends on its wavenumber. 
In the case of the GAM the damping rate increases with the GAM radial wavenumber (more precisely, with the radial wavenumber of the radial electric field). 
Since the GAM frequency depends on the temperature and safety factor, the GAM oscillates with different frequencies at different radial points in presence of the temperature gradient or magnetic shear. Distorting the GAM radial structure and creating higher radial wavenumbers, this process can strongly increase the GAM damping rate \cite{Zonca08, Chen95}.
A section \ref{PhaseMixing} of our paper is dedicated to the extension of previous works \cite{Biancalani17, Biancalani16}, which were done treating the electrons as adiabatic, and in circular flux surfaces, to the inclusion of kinetic electrons and realistic tokamak configurations. Finally, the last section of this paper is dedicated to the investigation of a realistic discharge of ASDEX Upgrade, described in Ref.~\cite{Conway08}. In Appendix \ref{appendix:ComGENE} we have shown a comparison between ORB5 and GENE for the case of non-flat temperature profile.
\section{Model}
The gyrokinetic simulations presented in this work have been performed with the code ORB5 \cite{Jolliet07, Bottino15}. ORB5 is a nonlinear gyrokinetic multi-species global particle-in-cell (PIC) code, which solves the Vlasov-Maxwell system in the electrostatic or electromagnetic limit, and has a capability of handling true MHD equilibrium for an axisymmetric toroidal plasma. 
The particle-in-cell method consists of coupling a particle-based algorithm for the Vlasov equation with a grid-based method for the computation of the self-consistent electromagnetic fields. 
Several physical models are available in ORB5, all of them derived from a systematic Hamiltonian theory \cite{Bottino15, Tronko16}  to provide exact energy and momentum conservation.  
In this work, only one ion species (deuterium) has been considered while the electrons are assumed to be drift-kinetic. This corresponds to the following gyrokinetic total Lagrangian:
\aeqns
L =&& \sum_{\rm{sp}}\int \d V \d W \left(\left(\frac{q}{c}\vecb{A}+p_\parallel\vecb{b}\right)\cdot\dot{\vecb{R}} + \frac{m c}{q}\mu \dot{\theta} - H_0-H_1\right)f \\
&&+ \int \d V \d W  H_2 f_{M,ions}-\int \d V \frac{B_\perp^2}{8\pi}.\nonumber
\eeqns
The velocity variables are the magnetic moment  $\mu\equiv(mv_\perp^2)/(2B)$, the canonical parallel momentum $p_\parallel$ and the gyroangle $\theta$.
The equilibrium magnetic field is $\vecb{B}=\nabla\times\vecb{A}$, $m$ and $q$ are the mass and charge of the particle species $sp$ and $c$ is the speed of light.
The volume element  of the velocity space is $\d W\equiv (2\pi)/m^2 B^*_\parallel \d p_\parallel \d \mu$ 
with $B^*_\parallel=\vecb{B^*}\cdot{\vecb{b}}$, $\vecb{b}=\vecb{B}/B$ and $\vecb{B^*}=\vecb{B}+(c/q)p_\parallel\nabla\times\vecb{b}$; $\d V$ denotes the volume element in physical space. 
Here $f$ is the distribution function for the species $sp$, while $f_{M,ions}$ is the equilibrium time independent distribution function of the ions.
In this system, only long wavelength electrostatic perturbation and magnetic perturbations perpendicular to the equilibrium magnetic field are considered.
Note that no second order term in the fields is retained for the electrons, this is equivalent to neglect the electron polarization density in the Polarization equation (drift-kinetic approximation, see Ref. \cite{Bottino15} for details). 
The first two terms in the total Lagrangian define the charged particles Lagrangian \cite{Sugama00}. 
The GK Hamiltonian in general depends on the electrostatic potentials $\Phi$ and on the
parallel component of the fluctuation magnetic potential $A_\parallel$. 
 The third term in the total Lagrangian is the electromagnetic field Lagrangian, in which the electric field component has been neglected (quasi-neutrality approximation, see \cite{Bottino15} for details).
In this work we used the following Hamiltonian:
\aeqn
H&=&H_0+H_1+H_2\\  
H_0&=&\frac{p_\parallel^2}{2m}+\mu B\nonumber\\
H_1&=& e(J_0\Phi - \frac{p_\parallel }{mc} J_0A_\parallel)\nonumber\\ 
H_2 &=& - \frac{q^2}{2mc^2}(J_0A_\parallel)^2+\frac{mc^2}{2B^2}|\nabla_\perp \Phi|^2\nonumber
\eeqn
the gyroaveraging (Hermitian) operator $J_0$, applied to an arbitrary function $\psi$ in configuration space, is defined by
\aeq
(J_0\psi) ( \mathbf{R},\mu) = \frac{1}{2\pi}\int_0^{2\pi} \psi( \mathbf{R}+\boldsymbol{\rho}(\alpha)) \,d \alpha, 
\eeq
where $\boldsymbol{\rho}$ is the vector going from the guiding center position to the particle position.
In this work we have assumed $J_0=1$ for the electrons (drift-kinetic approximation).
The gyrokinetic equations for the particle distribution function and the GK
field equations can be derived from the GK Lagrangian using variational
principles. In summary, the GK model used in the following is:\\
$\bullet$ gyrokinetic full-f Vlasov equation for the ions
\aeqns
&&\fpar{f_i}{t}+\dot{\vecb R_i}\cdot \nabla f_i+\dot{p_{\parallel,i}}\fpar{f_i}{p_{\parallel,i}}=0,\label{GK1}\\
&&\dot{\vecb R_i}=\left(\frac{p_{\parallel,i}}{m_i}-\frac{Z_ie}{m_ic}J_0A_\parallel \right)\frac{\vecb{B_i^*}}{B^*_{\parallel,i}}+\frac{c}{Z_ieB^*_{\parallel,i}}\vecb{b}\times\left[\mu_i
  \nabla B + Z_ie \nabla J_0 \Psi_i \right],\label{GK2}\\
&&\dot{p_{\parallel,i}}=-\frac{\vecb{B_i^*}}{B^*_{\parallel,i}}\cdot\left[\mu_i \nabla B + Z_ie
  \nabla J_0\Psi_i \right],\label{GK3}
\eeqns
$\bullet$ drift-kinetic full-f Vlasov equation for the electrons:
\aeqns
&&\fpar{f_e}{t}+\dot{\vecb R_e}\cdot \nabla f_e+\dot{p_{\parallel,e}}\fpar{f_e}{p_{\parallel,e}}=0,\label{GK1e}\\
&&\dot{\vecb R_e}=\left(\frac{p_{\parallel,e}}{m_e}+\frac{e}{m_ec}A_\parallel \right)\frac{\vecb{B_e^*}}{B^*_{\parallel,e}}-\frac{c}{eB^*_{\parallel,e}}\vecb{b}\times\left[\mu_e
  \nabla B - e \nabla \Psi_e \right],\label{GK2e}\\
&&\dot{p_{\parallel,e}}=-\frac{\vecb{B_e^*}}{B^*_{\parallel,e}}\cdot\left[\mu_e \nabla B -e
  \nabla \Psi_e \right],\label{GK3e}
\eeqns
having introduced the generalized potential
\aeq
\Psi \equiv \Phi - \frac{p_{\parallel,sp} }{m{_{sp}}c} A_\parallel.
\eeq
$\bullet$ Linear polarization equation in the long wave-length limit (and drift-kinetic electrons):
\aeqns
&&\int \d W_i  Z_ieJ_0 f_i-\int \d W_e  e f_e=-\nabla \cdot \left(\frac{n_0 m_ic^2}{B^2} \nabla_\perp \Phi \right)\label{GK4}
\eeqns
$\bullet$ Linear Amp\`ere's law:
\aeqns
\int \d W_i   \frac{4\pi Z_ie }{m_i c} p_{\parallel,i} J_0 f_i &-& \int \d W_e  \frac{4\pi e }{m_e c} p_{\parallel,e} f_e = \\
&&\frac{1}{d_e^2}A_\parallel+\frac{1}{d_i^2}A_\parallel-\nabla_\perp^2 A_\parallel
-\nabla\cdot\frac{\beta_{i}}{4}\nabla_\perp A_\parallel\label{GK5}
\eeqns
where $n_0$ is the density associated with the equilibrium Maxwellian $f_M$ and  $\beta_{sp}=(4\pi n_{sp} T_{sp})/B^2$. The skin depth is defined by $d^{-2}_{sp}=\beta_{sp}/\rho^2_{sp}$, where $\rho^2_{sp} = T_{sp}m_{sp}c^2/q^2_{sp}B^2$, and it appears on the right-hand-side of the Amp\'ere's law
because of the choice of the velocity space variables $(p_\parallel,\mu)$ instead of the usual $(v_\parallel,\mu)$.
The indexes $i$ and $e$ indicate ions and electrons respectively and $q_i=Z_ie$ is the ion charge while $q_e=-e$ is the electron charge.
Despite all the approximations made, this model is highly physically relevant and it
can be used to describe not only the GAM and ZF dynamics, but also a large class of micro-instabilities excited by the density
and temperature gradients, like ion temperature gradient (ITG) driven modes,
trapped electron modes (TEM) or kinetic ballooning modes (KBM). It also contained the reduced MHD model as a subset (see, among other, Ref. \cite{Miyato09}).

According to the PIC method the particle distribution function is discretized with macroparticles, known as markers. 
The motion of the markers is calculated using the equations of motions of the gyrokinetic model while the 
electromagnetic fields are evolved on a spatial grid using the two field equations. The charge and current density, that are necessary to solve the field equations, 
are calculated by projecting the marker weights on a spatial grid. After that, the fields are calculated using a finite elements method.

The code is based on a straight-field-line coordinate system $(s, \chi, \phi)$. Here, radial coordinate is $s = \sqrt{\psi/\psi_{edge}}$ (where $\psi$ is the poloidal flux), $\chi = \yfrac{1}{q(s)} \int_0^\Theta \yfrac{\yB \cdot \nabla \phi}{\yB \cdot \nabla \Theta_1} d\Theta_1$  is the straight-field-line coordinate (where $\Theta$ and $q$ are the poloidal angle and the safety factor respectively) and $\phi$ is a toroidal angle. 
Two different kinds of magnetic equilibria are implemented: analytical equilibria with circular concentric magnetic surfaces and ideal MHD realistic equilibria. For the latter case, the ORB5 code is coupled with the CHEASE code \cite{Lutjens96}, which solves the Grad-Shafranov equation with a fixed plasma boundary.
\section{Frequency and Landau damping}
\subsection{Equilibrium and simulations parameters}
\label{PARS_WOMIX}
Linear electromangetic gyrokinetic collisionless simulations with drift-kinetic electrons and a realistic electron - ion (deuterium) mass ratio $m_e / m_i = 2.5\cdot10^{-4}$ have been performed. 
Electrostatic simulations with kinetic electrons are, in principle, faster than electromagnetic simulations, due to the smaller number of equations to be solved. 
Nevertheless, a high frequency oscillation, called the $\omega_H$-mode~\cite{Lee87}, is observed to be often numerically unstable.
To decrease the level of the high-frequency oscillations, electromagnetic simulations in the small-$\beta_e$ ($\beta_e = 10^{-5}$) limit have been performed instead of the electrostatic ones.
MHD equilibria of the circular and elongated plasma have been calculated with an external code CHEASE \cite{Lutjens96}.
Simulations have been carried out with a flat density profile, which have been shown to not impact the GAM frequency and damping rate in linear simulations (see Appendix \ref{appendix:convergence}).
To focus on the Landau damping in the absence of the phase mixing effect, flat temperature profile has been considered in simulations used for the results of this section.
Since the safety factor profiles have been taken from the CHEASE, there is a magnetic shear, that also causes the phase mixing, but its influence on the GAM damping rate is much smaller in comparison to the temperature gradient effect.

Plasma parameters have been taken close to the ASDEX Upgrade parameters near the plasma edge \cite{Conway08}: the major radius $R = 1.65$ m, the minor radius $a = 0.5$ m (inverse aspect ratio is $\epsilon = 0.303$), the magnetic field on the axis $B = 2$ T.
Since in the CHEASE code the plasma elongation is defined at the edge and changes gradually to the plasma center, the GAM frequency and damping rate have been measured at the same radial position $s_0 = 0.90$ to perform more accurate scan on the elongation. 
The temperature has been taken to be $T_i = T_e = 70$ eV. 
It means that $c_s = \sqrt{q_e T_e / 2 m_p} = 5.8 \cdot 10^{3}$ m/s, $\rho_s = c_s / \omega_{ci} = 6.1\cdot10^{-4}$ m and $\rho^* = \rho_s / a = 1.2\cdot10^{-3}$. 
Here, $\omega_{ci} = q_e B/2m_p$ ($\omega_{ci}/2\pi = 15.2$ MHz) is the ion gyro-frequency, $q_e$ is an electron charge, and $m_p$ is a proton mass. 

To weaken the constraint on the space step and to reduce the effect of the charge accumulation at the edge of the numerical work box, we have simulated only a ring from $s_1 = 0.85$ to $s_2 = 0.95$ in a poloidal cross section with the Dirichlet condition for the potential $\phi$ on inner boundaries ($\phi(s_1) = \phi(s_2) = 0$).

A typical simulation has the following parameters. 
Number of nodes in radial direction is taken to be $n_s = 256$, in toroidal directions $n_{\phi} = 4$ and along the straight-field-line coordinate $\chi$ the number of nodes is $n_{\chi} = 64$. 
Time step is $dt \cdot \omega_{ci} = 2$. 
The GAM damping rate and frequency have been calculated (see Appendix \ref{appendix:convergence}) for different GAM radial wavenumbers $k = k_r\rho_i \in [0.054, 0.377]$ (where $\rho_i = \sqrt{2}v_{Ti}/\omega_{ci} = 8.57 \cdot 10^{-4}$ m is the ion Larmor radius of the deuterium and $v_{Ti} = \sqrt{q_eT_i/2m_p}$ is the thermal speed), the safety factor $q \in [3.5, 5.0]$ at $s_0 = 0.90$ and the plasma elongation $e \in [1.0, 1.6]$ at the edge. This is the regime where GAMs are typically observed in tokamak plasmas (see, for example, Ref. \cite{Conway08}). To simulate the GAM dynamics, the ORB5  simulations have been initialized by introducing an axisimmetric density perturbation designed to produce an initial electric potential field of the form $\sim \sin(ks)$, where $s \in [s_1, s_2]$ (as in the so-called Rosenbluth-Hinton test \cite{Rosenbluth98}). All toroidal modes $n \neq 0$ and poloidal modes $|m| > 10$ have been filtered out. To study the GAM dynamics, the frequency and damping rate of the poloidally averaged radial electric field have been calculated.
\subsection{Results of gyrokinetic simulations}
\label{GK_RES}

\begin{figure}[b!]
\begin{subfigure}[t]
{0.49\textwidth} 
\includegraphics[width=\textwidth]{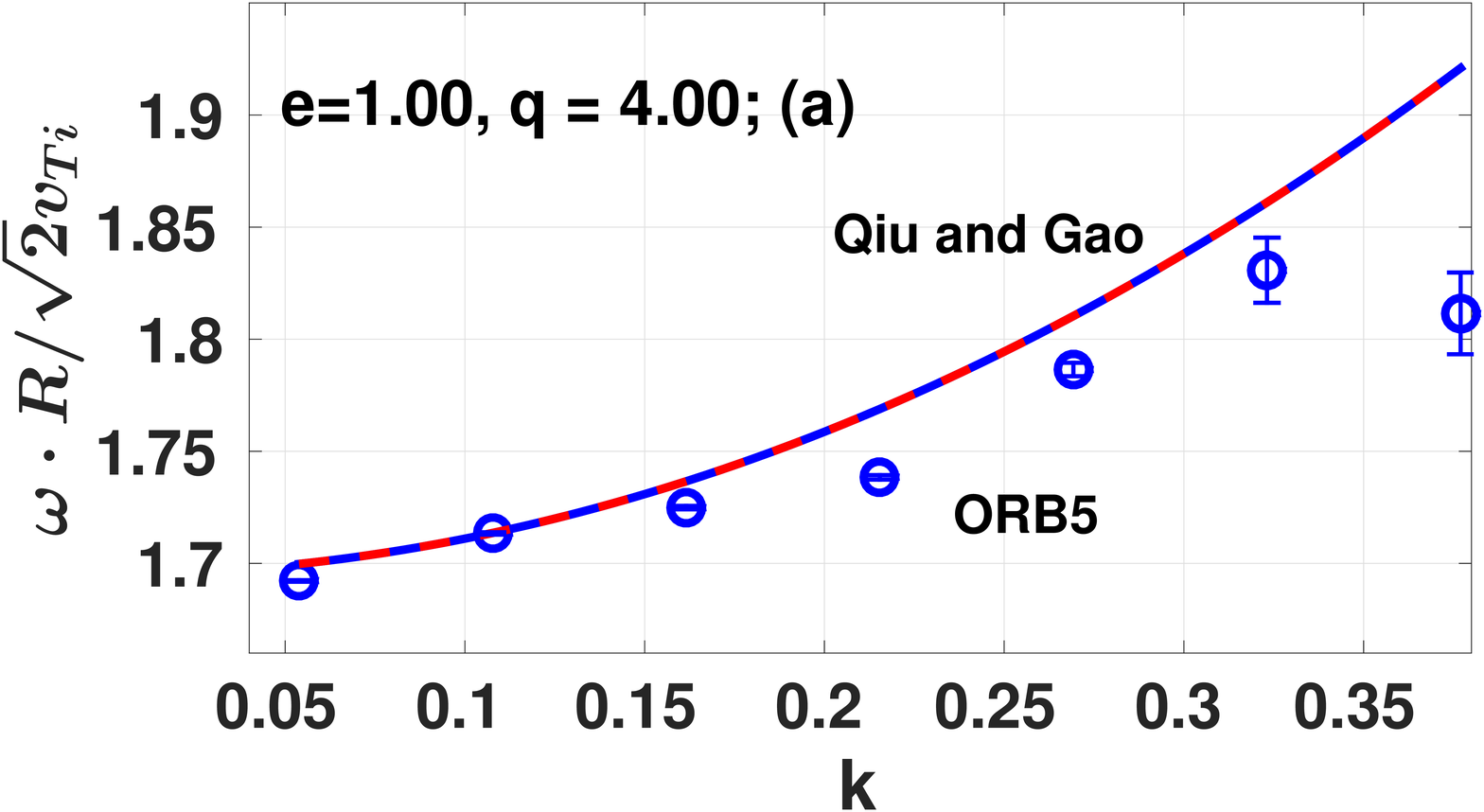}
\end{subfigure}
\begin{subfigure}[t]
{0.49\textwidth} 
\includegraphics[width=\textwidth]{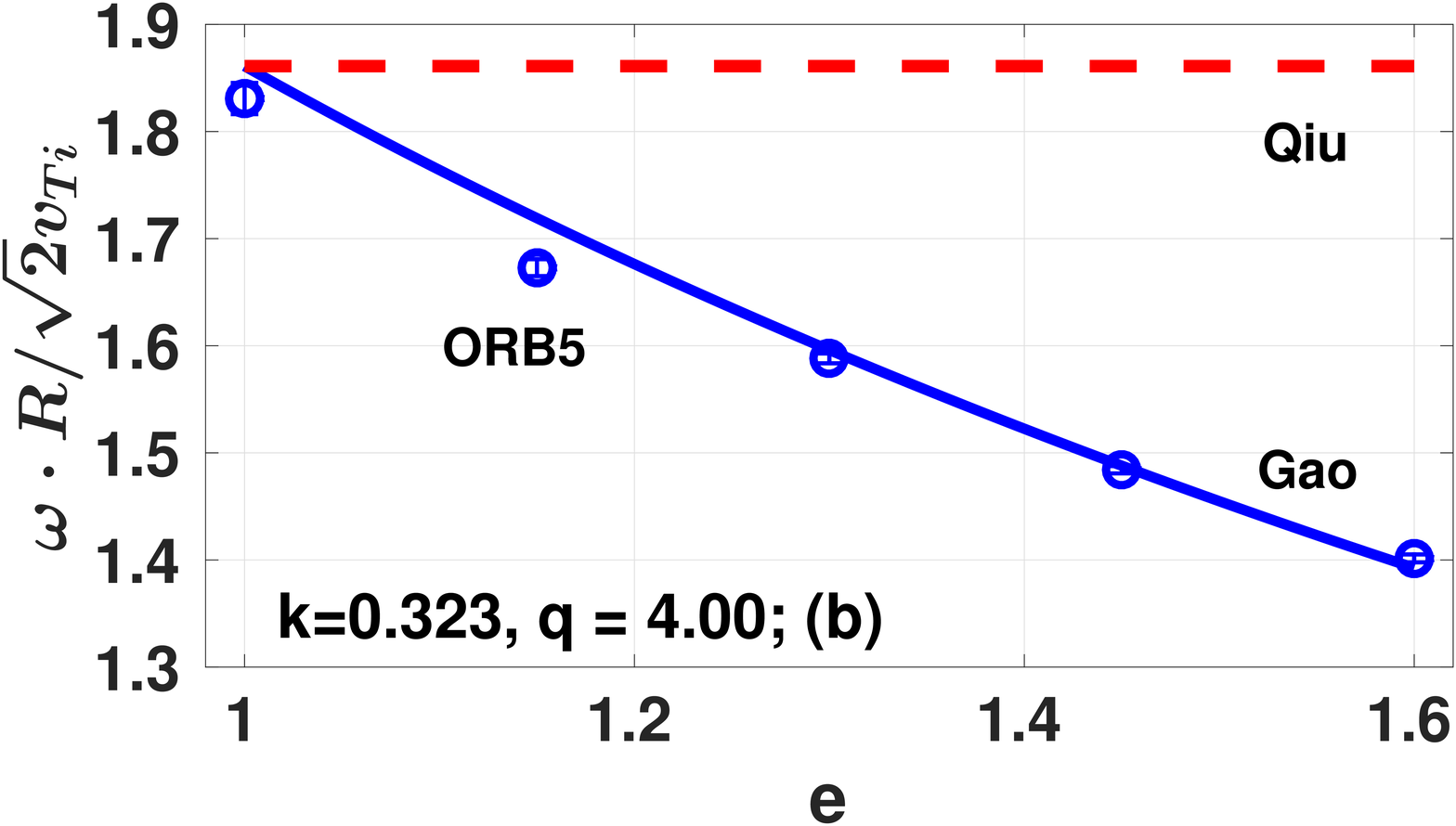}
\end{subfigure}
\begin{subfigure}[t]
{0.49\textwidth} 
\includegraphics[width=\textwidth]{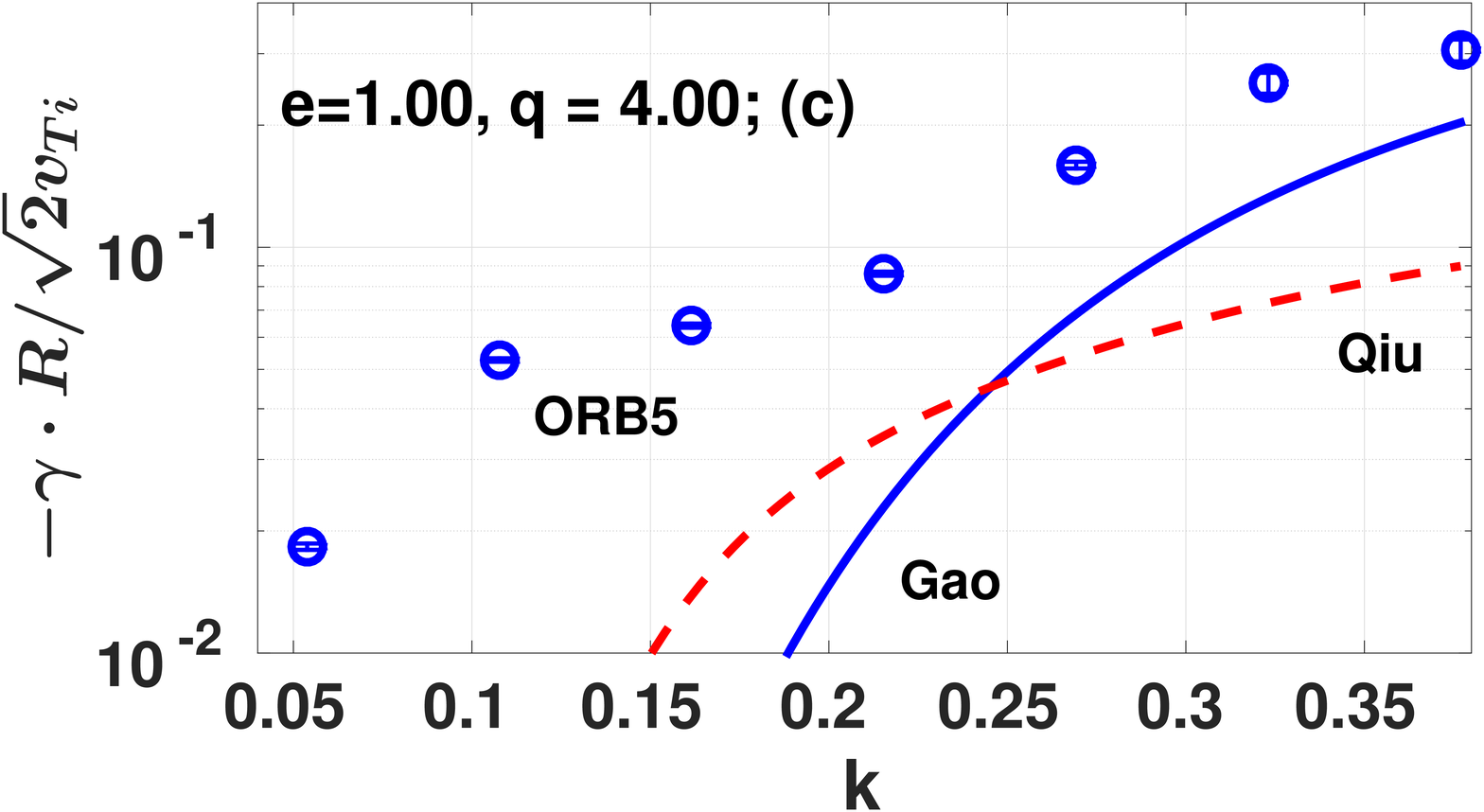}
\end{subfigure}
\begin{subfigure}[t]
{0.49\textwidth} 
\includegraphics[width=\textwidth]{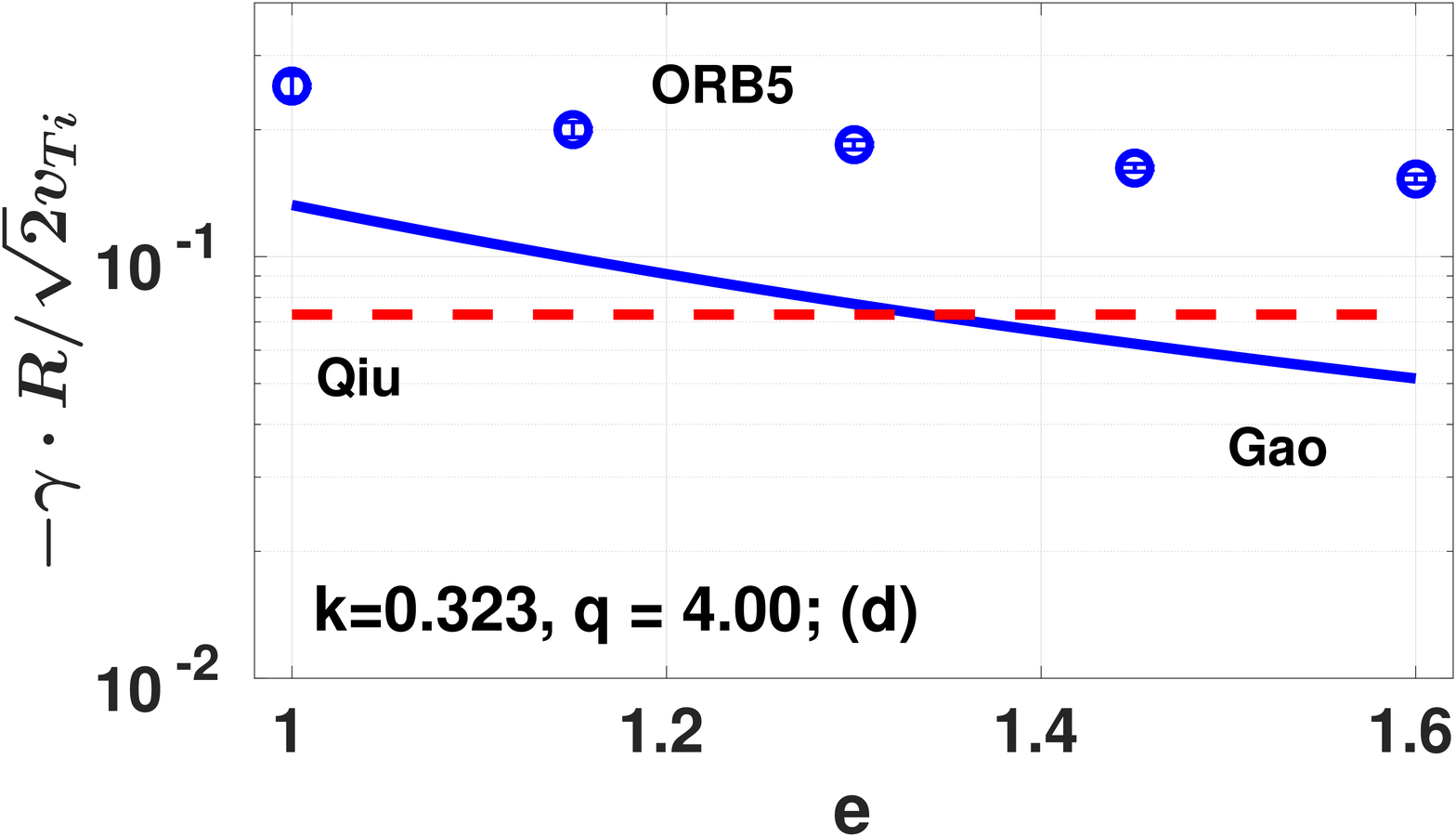}
\end{subfigure}
\caption{\label{fig:COMP} Comparison of the results from linear gyrokinetic simulations (blue dotes), performed with kinetic electrons, with the analytical theories Gao 2010 \cite{Gao10} (solid blue line) and Qiu 2009 \cite{Qiu09} (dashed red line), derived with adiabatic electrons. Here, $k = k_r \rho_i$.}
\end{figure}
In Fig. \ref{fig:COMP}, a comparison of the GAM frequencies and damping rate obtained from numerical simulations with two analytical theories of Qiu 2009 \cite{Qiu09} and Gao 2010 \cite{Gao10} is  shown.
A good agreement between numerical results and analytical predictions of the GAM frequency has been found. 
Nevertheless, the GAM damping rate, obtained from the theories, derived using adiabatic electrons, is smaller in comparison to numerical simulations with kinetic electrons, and the divergence increases for smaller values of the GAM radial wavenumber. Moreover, since the frequency stops increasing in the domain of higher wavenumbers (subplot $a$ of Fig. \ref{fig:COMP}), a divergence between numerical results and analytical theories is observed. The same effect was observed in Ref. \cite{Singh17}, where the GAMs were studied using drift reduced Braginskii equations.

The Gao 2010 theory describes the GAM dependence on the plasma elongation and it is in a good agreement with numerical results for the frequency. Although the Gao 2010 theory provides considerably smaller damping rate, it seems to give similar trend of the damping coefficient with the plasma elongation, i.e., the damping rate is weakened by the elongation. The Gao 2010 theory was derived in the large orbit drift width limit, where the dominant damping mechanism is the resonance $\omega \sim \omega_d$ (here, $\omega_d = \yb{k_r}\cdot\yb{v_d}$ is a magnetic drift frequency, $\yb{k_r}$ is a wave vector of the zonal potential in the radial direction and $\yb{v_d}$ is a magnetic drift velocity) \cite{Gao11}. 
As explained in Ref. \cite{Gao10}, the GAM frequency decreases with the elongation less rapidly than the drift frequency. To satisfy the resonance $\omega \sim \omega_d$ particles have to have higher drift velocities, which involves fewer particles in the wave-particle interaction and, as a result, the GAM damping rate decreases.
\begin{figure}[b!]
\begin{subfigure}[t] 
{0.49\textwidth} 
\includegraphics[width=\textwidth]{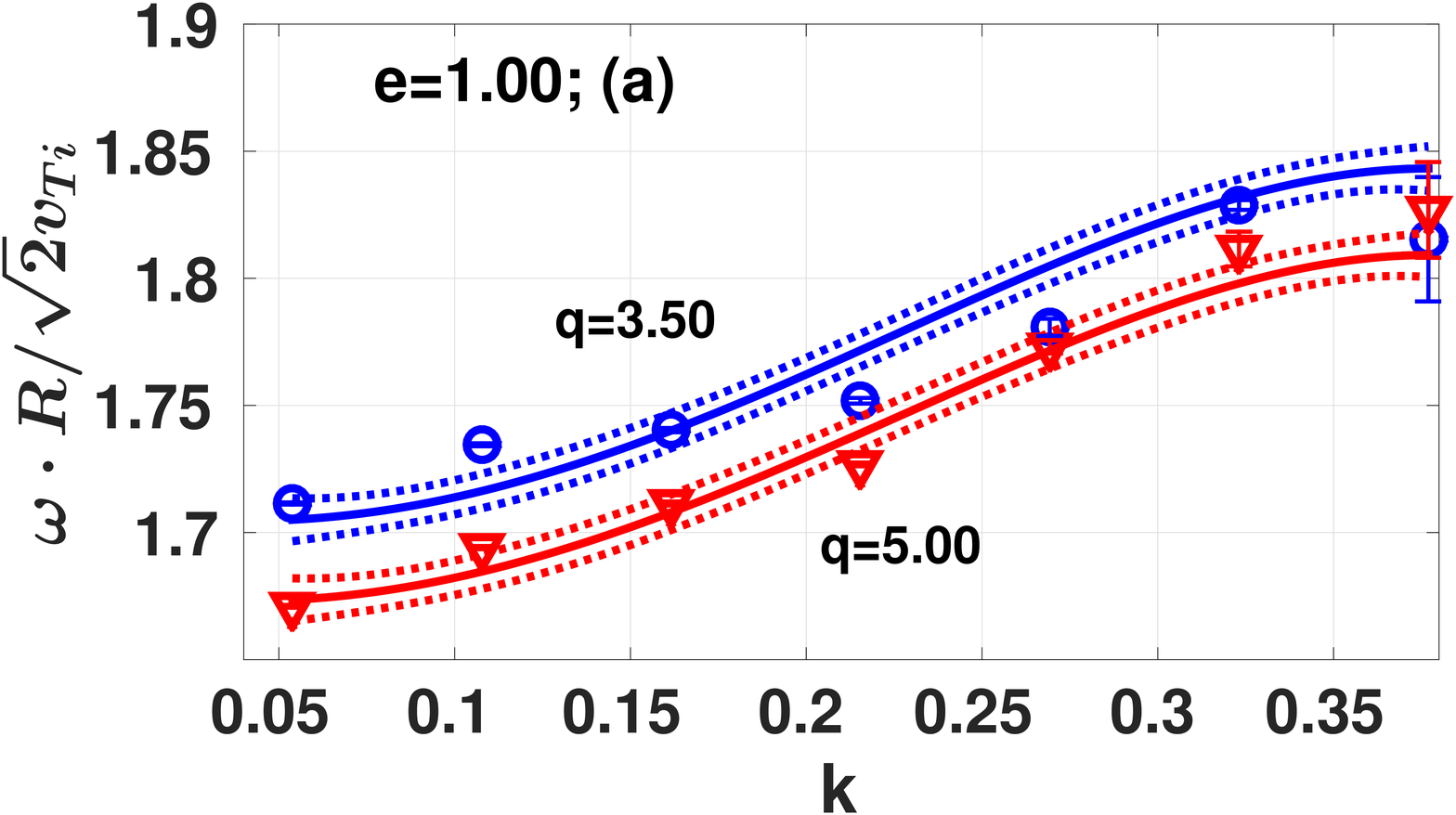}
\end{subfigure}
\begin{subfigure}[t]
{0.49\textwidth} 
\includegraphics[width=\textwidth]{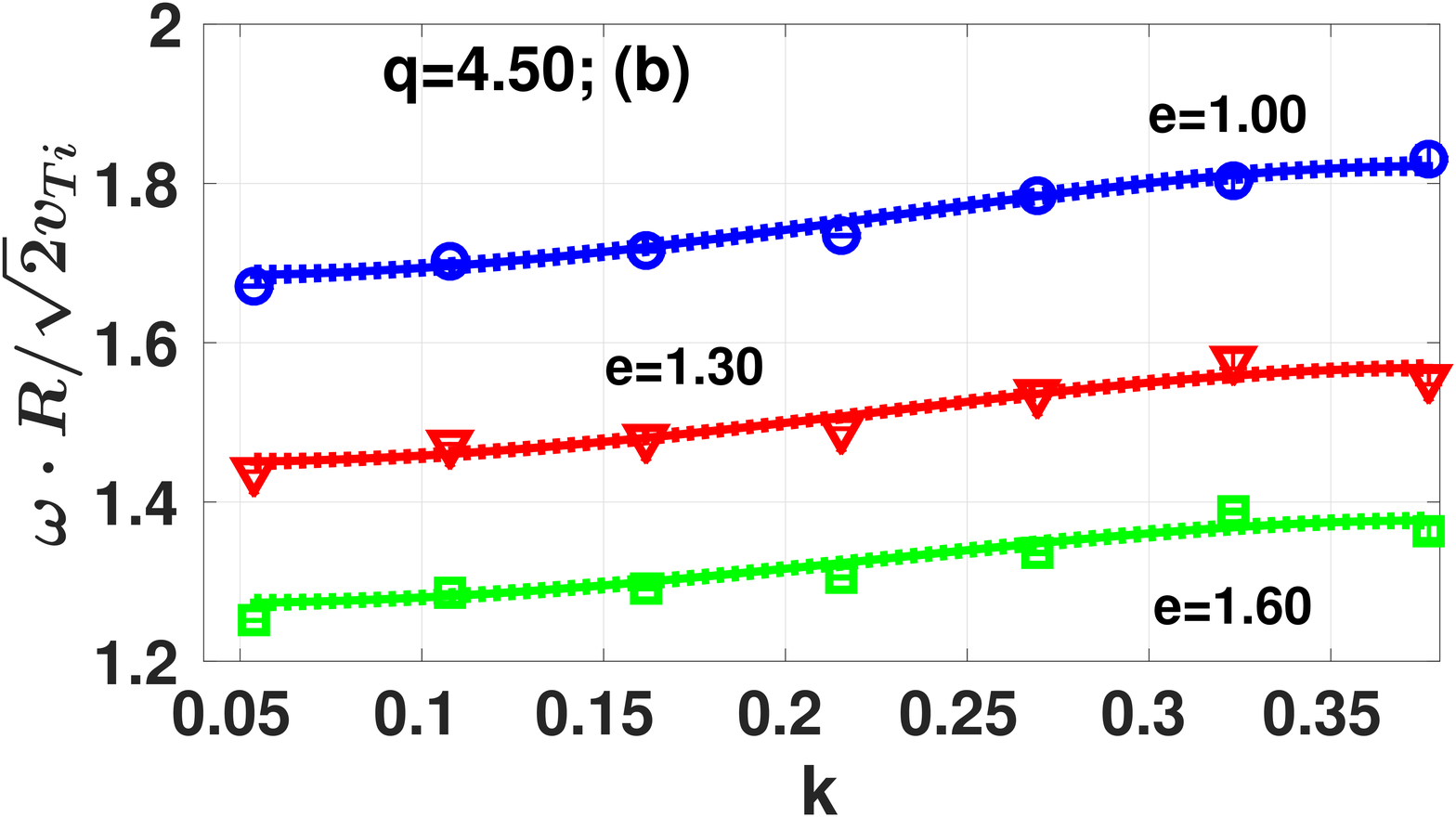}
\end{subfigure} 
\begin{subfigure}[t]  
{0.49\textwidth} 
\includegraphics[width=\textwidth]{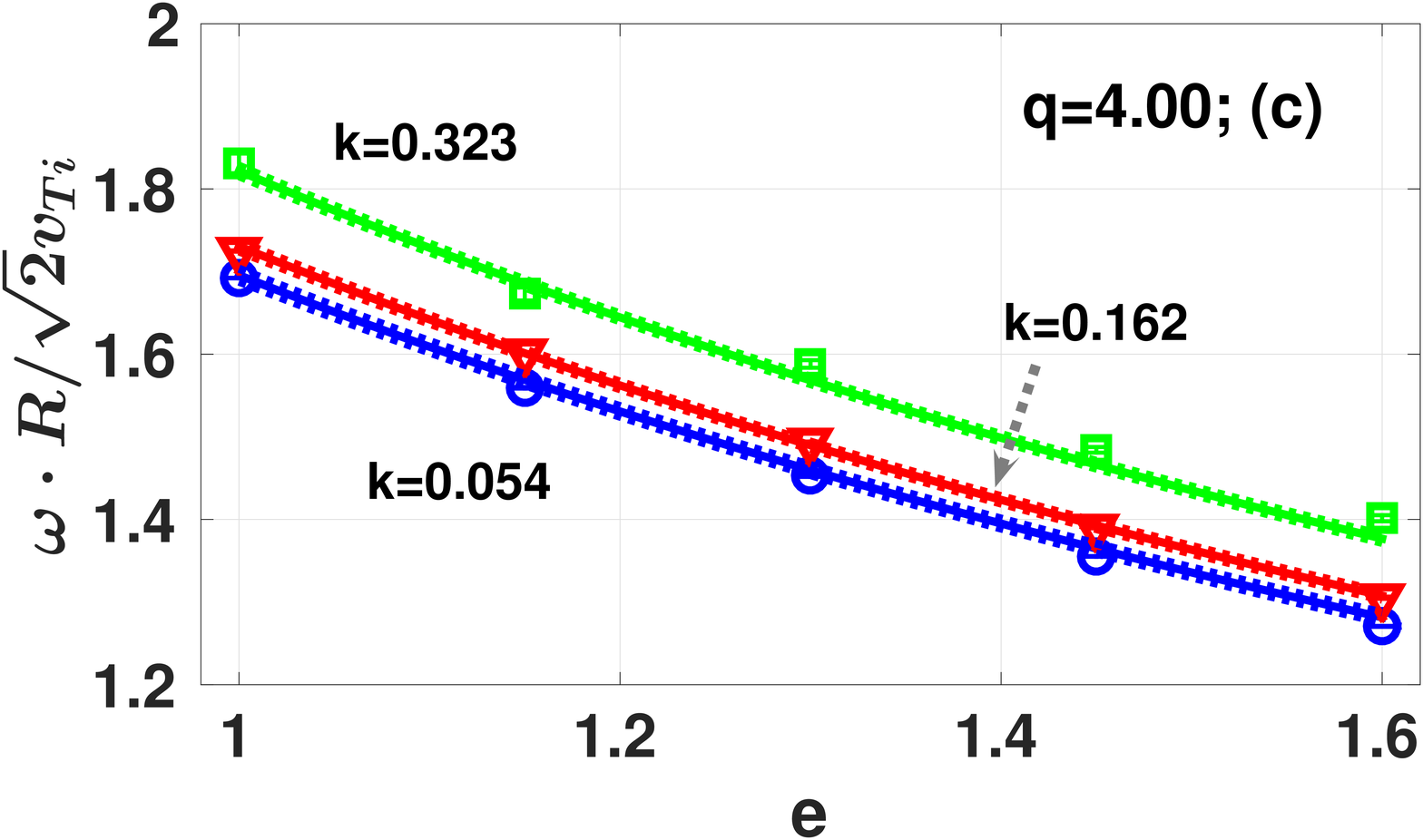}
\end{subfigure}
\begin{subfigure}[t]
{0.49\textwidth} 
\includegraphics[width=\textwidth]{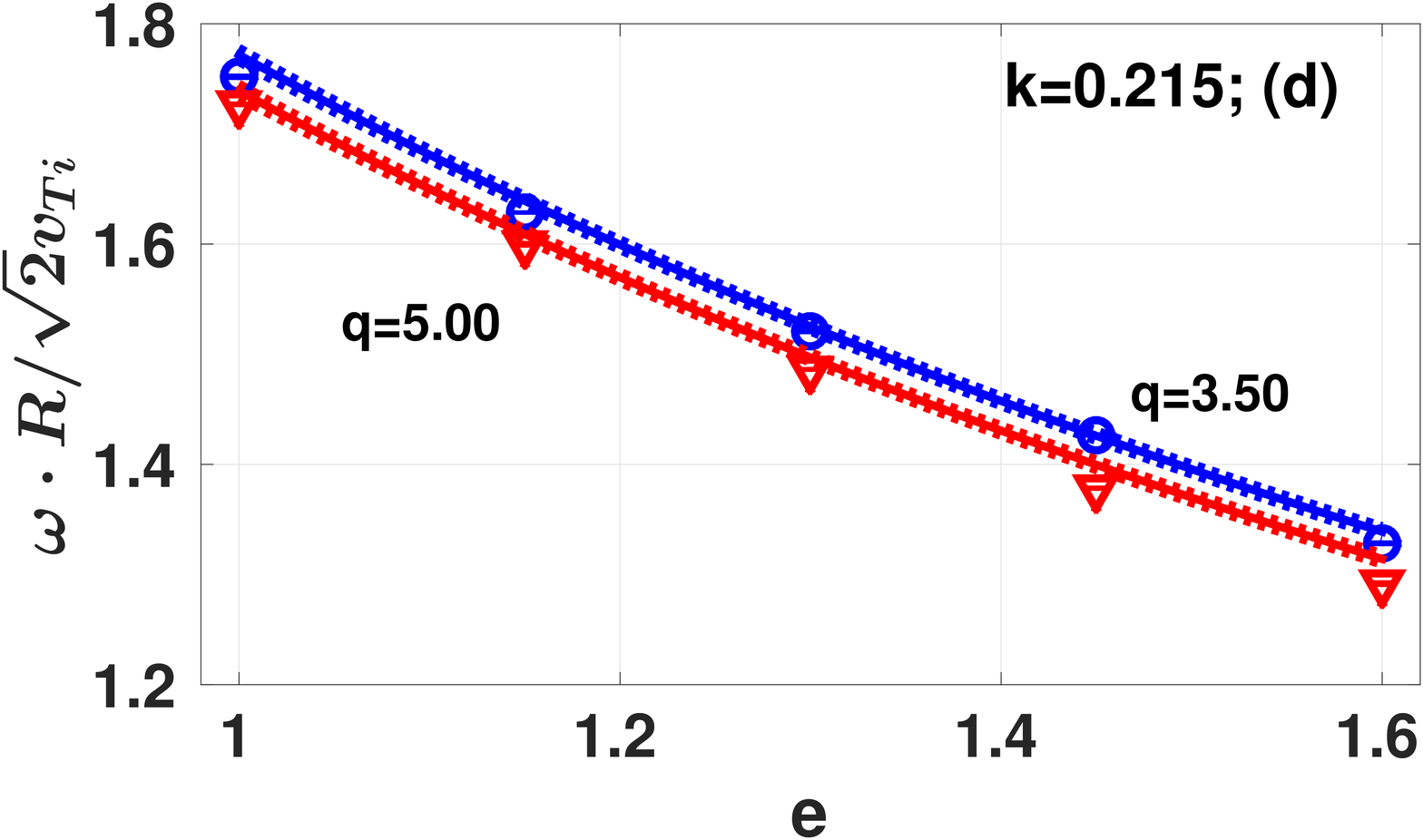}
\end{subfigure}
\begin{subfigure}[t]  
{0.49\textwidth} 
\includegraphics[width=\textwidth]{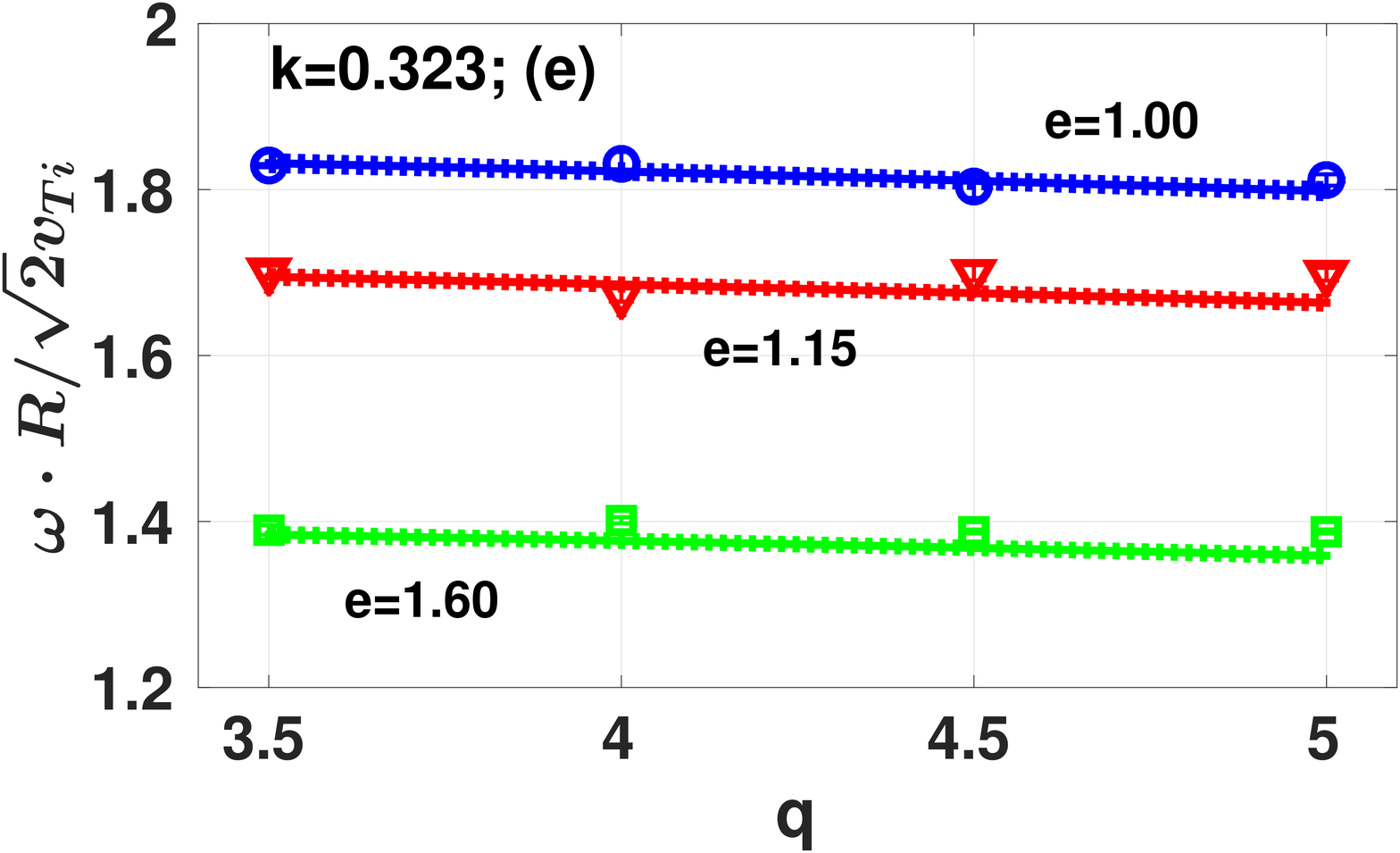}
\end{subfigure}
\begin{subfigure}[t]  
{0.49\textwidth} 
\includegraphics[width=\textwidth]{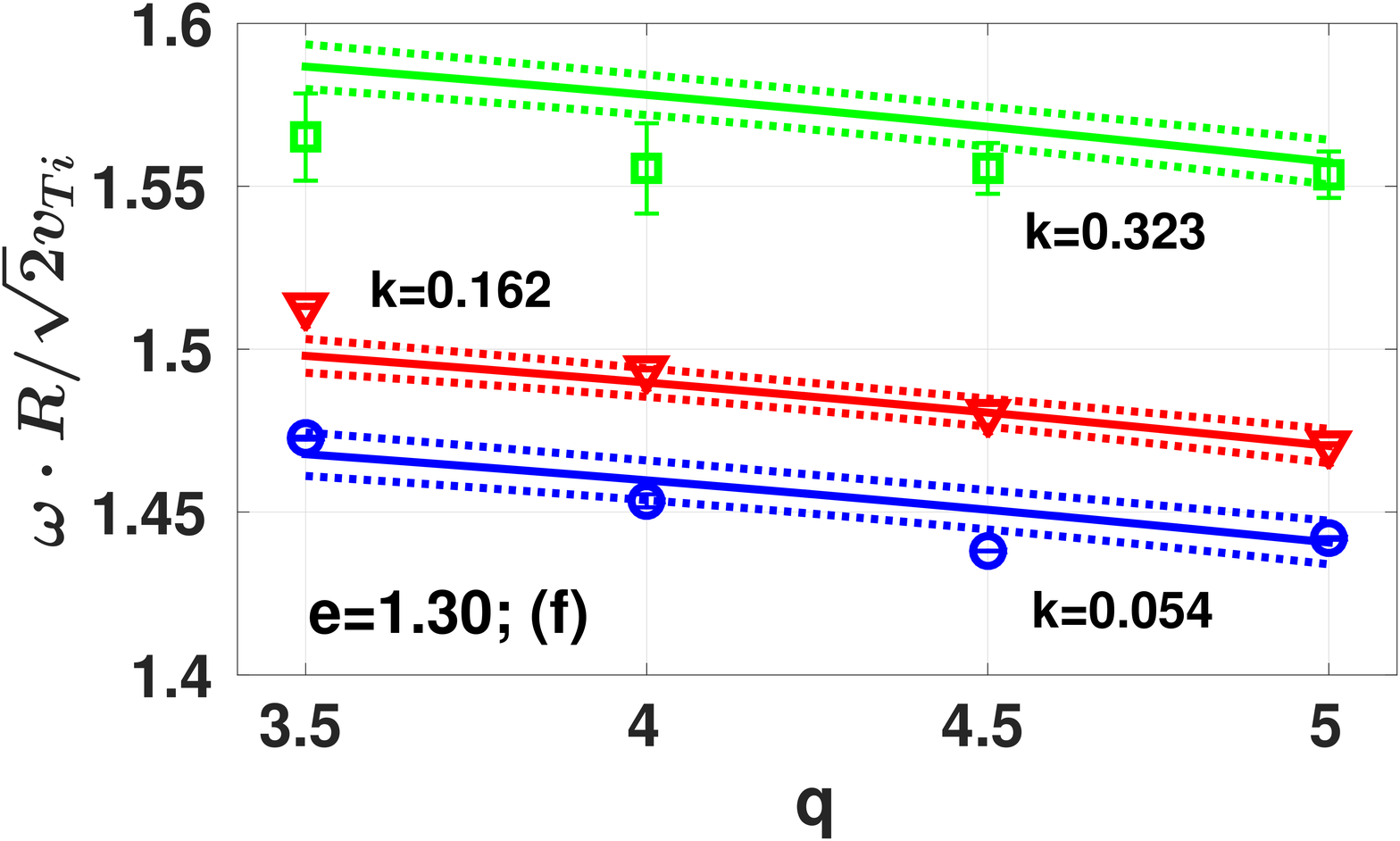}
\end{subfigure}
\caption{\label{fig:FIT_w} Comparison between numerically simulated values (dots, traingles and squares) of the GAM frequency and values obtained using the interpolating expression provided in Eq. (\ref{formula:FIT_w}) (solid lines). Dotted lines indicate 95\% confidence bounds of the fitting.}
\end{figure}
\subsection{Interpolating formulae}
\label{ch:InterpExp}
To provide a scaling of the GAM frequency and damping rate, corresponding interpolating expressions have been fitted to the results of the gyrokinetic simulations described in Sec. \ref{GK_RES}. For consistency, the regime has been chosen for the GAM wavenumbers $k = k_r \rho_i$ in the range $[0.054, 0.377]$, safety factor $q \in [3.5, 5.0]$ and plasma elongation $e \in [1.0, 1.6]$. 
\begin{figure}[b!]
\begin{subfigure}[t] 
{0.49\textwidth} 
\includegraphics[width=\textwidth]{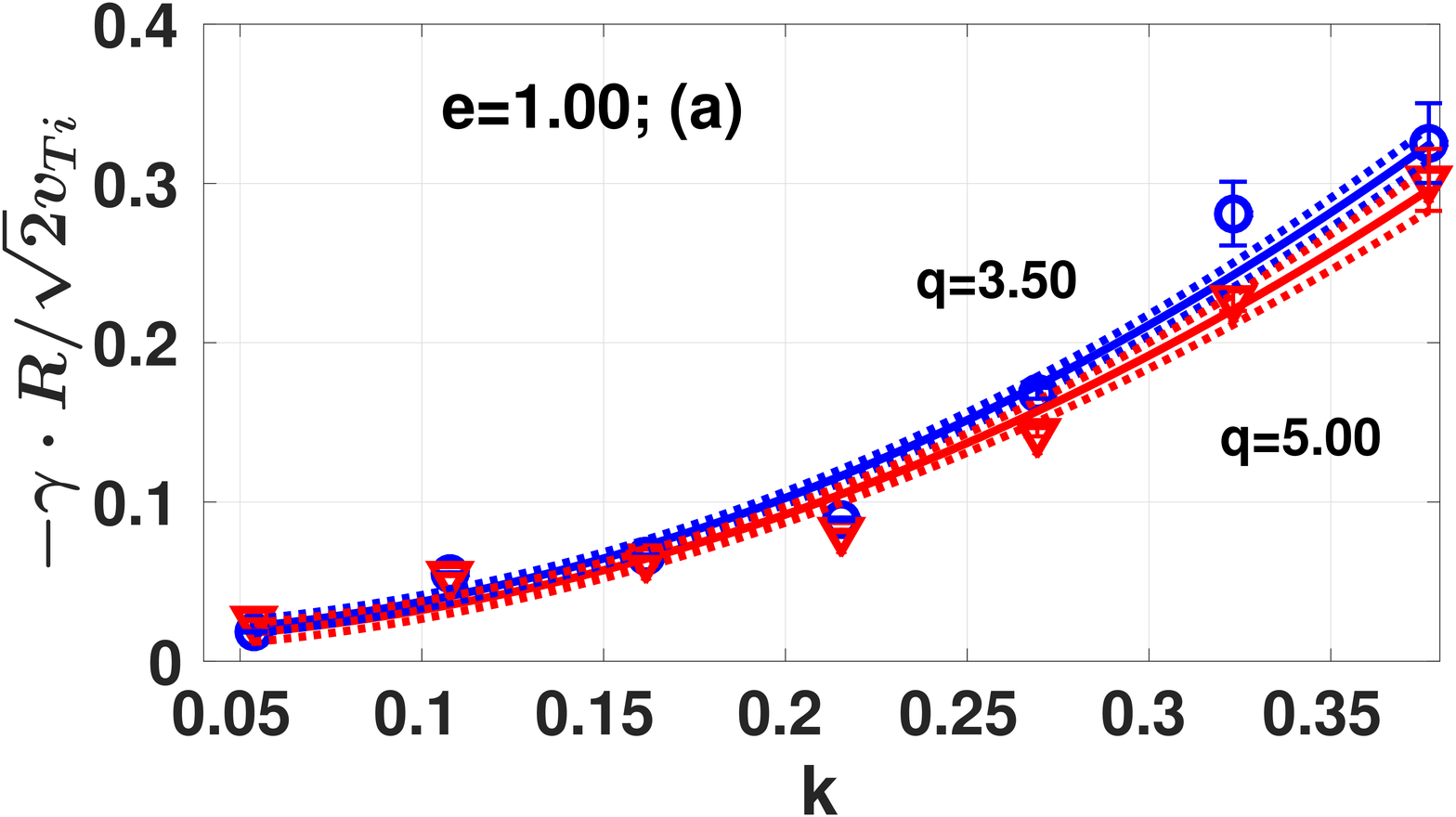}
\end{subfigure}
\begin{subfigure}[t]
{0.49\textwidth} 
\includegraphics[width=\textwidth]{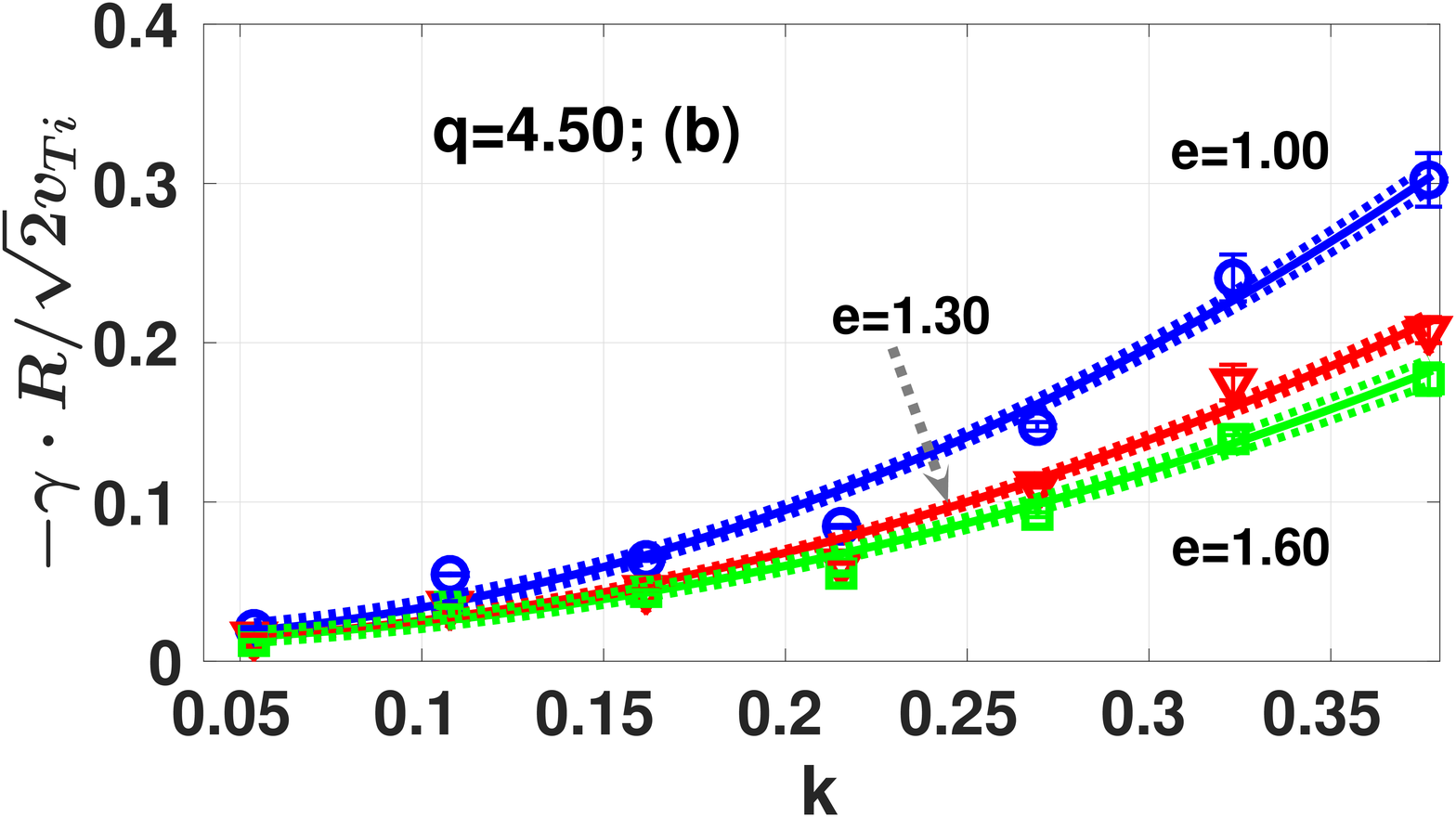}
\end{subfigure} 
\begin{subfigure}[t]  
{0.49\textwidth} 
\includegraphics[width=\textwidth]{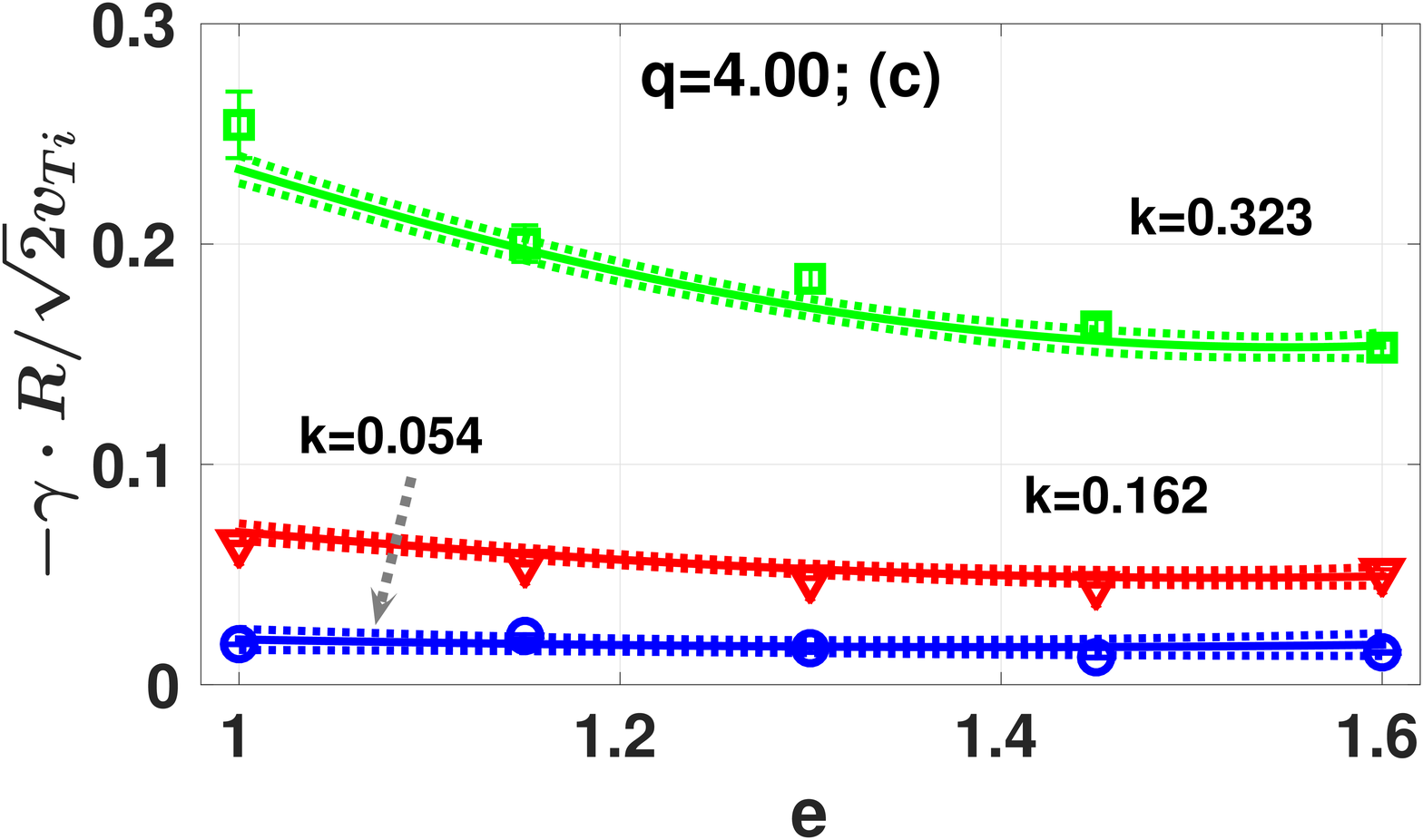}
\end{subfigure}
\begin{subfigure}[t]
{0.49\textwidth} 
\includegraphics[width=\textwidth]{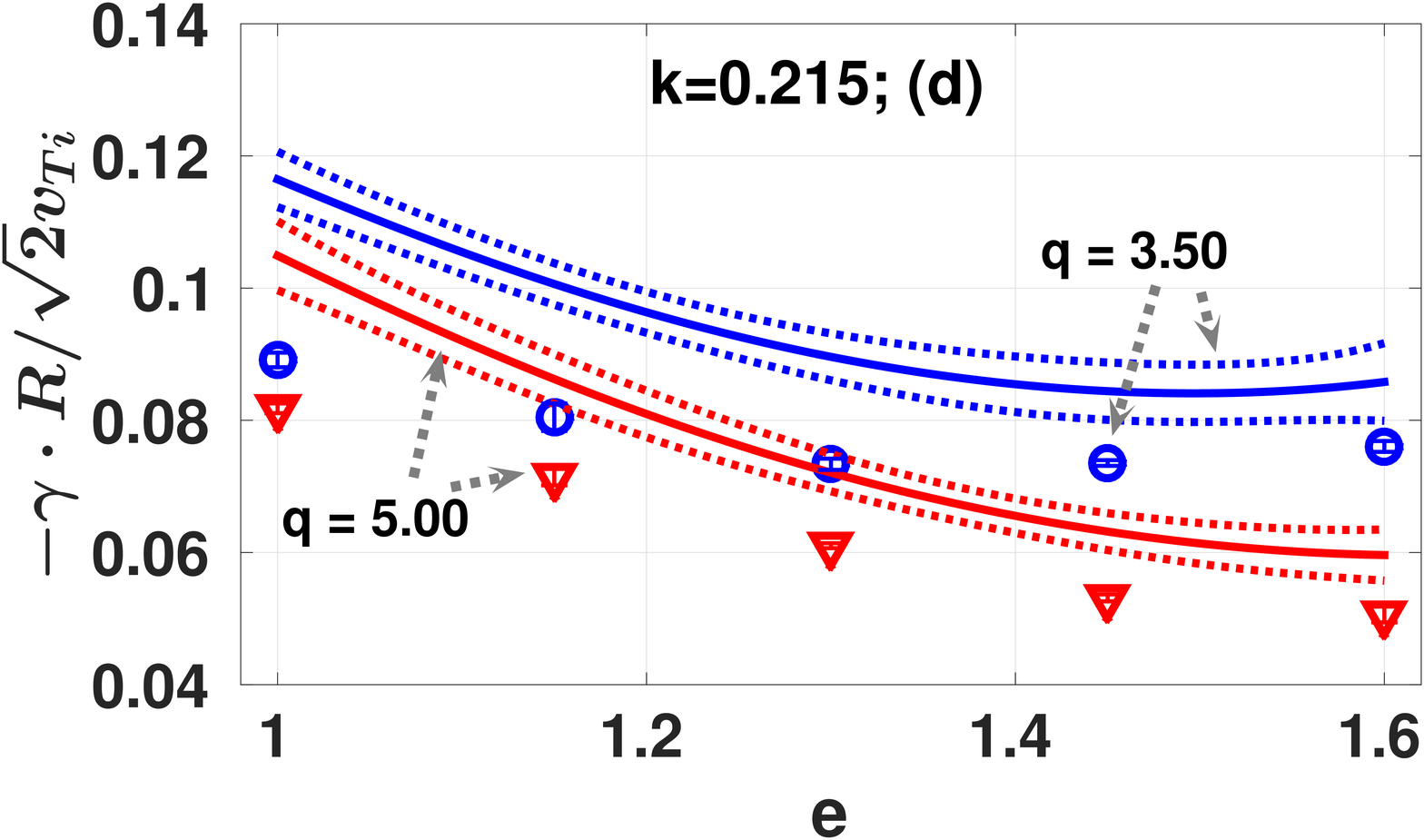}
\end{subfigure}
\begin{subfigure}[t]  
{0.49\textwidth} 
\includegraphics[width=\textwidth]{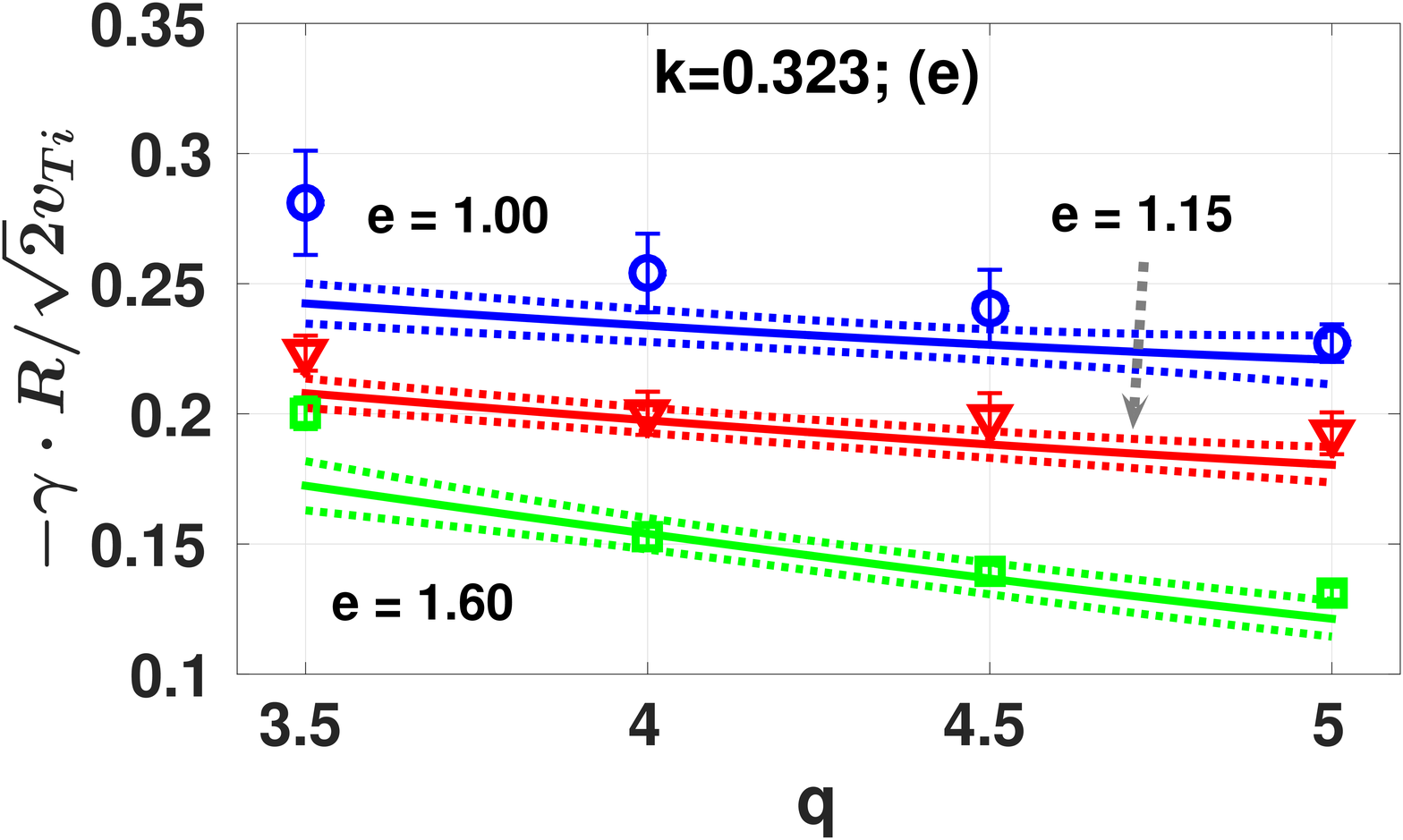}
\end{subfigure}
\begin{subfigure}[t]
{0.49\textwidth} 
\includegraphics[width=\textwidth]{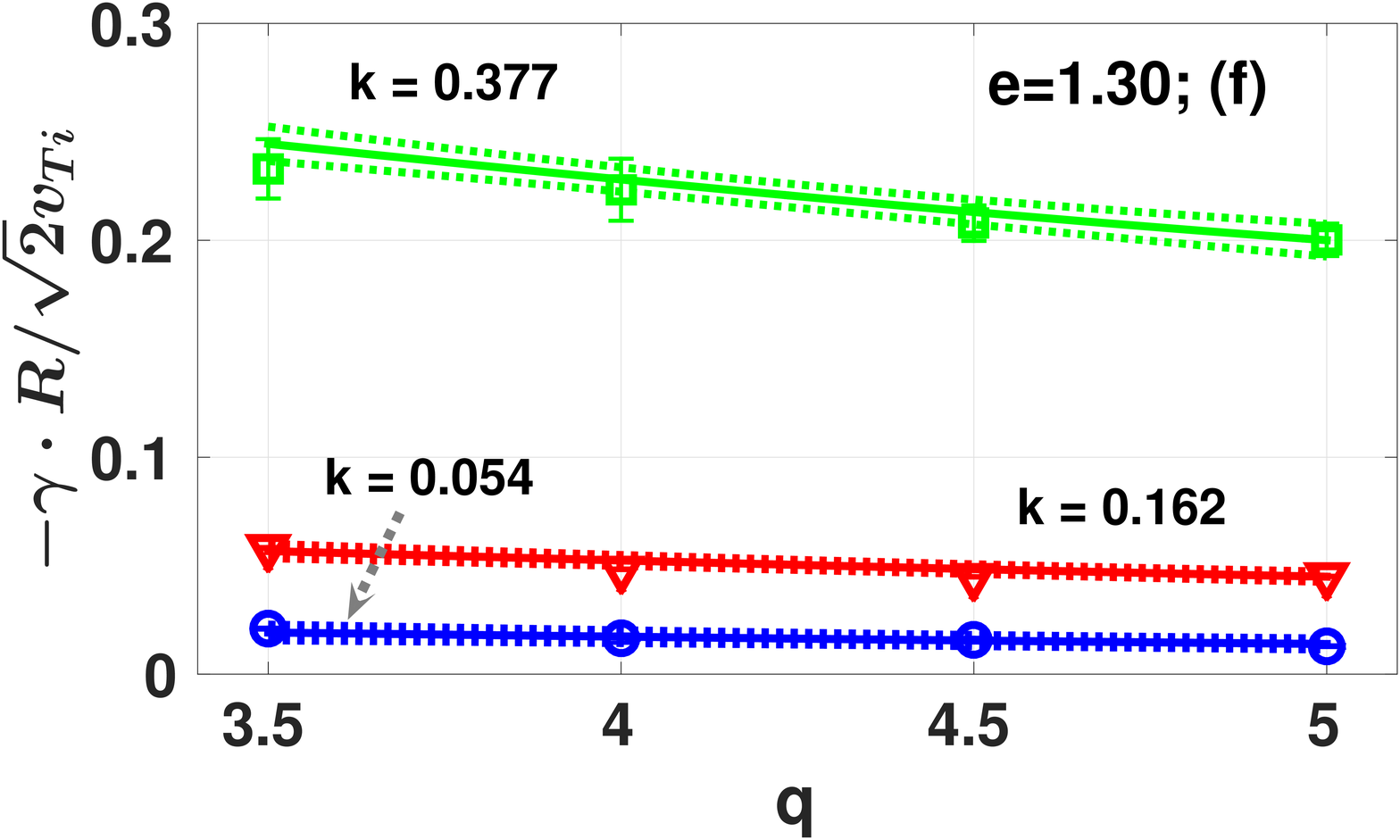}
\end{subfigure}
\caption{\label{fig:FIT_g} Comparison between numerically simulated values (dots, triangles or squares) of the GAM damping rate and values obtained by using the interpolating expression provided in Eq. (\ref{formula:FIT_g}) (solid lines). Dotted lines indicate 95\% confidence bounds of the fitting.}
\end{figure}

To derive an interpolating expression for the frequency several assumptions have been used. 
The experimentally obtained dependence \cite{Conway08} on the plasma elongation $1 / (1 + e)$ has been slightly modified to $1 / (1 + g_6 e)$, where $g_6$ is an adjustable coefficient.
The dependence on the safety factor has been taken in the form $\exp(-g_5 q^2)$. In fact, the $q$-dependence in a form of $\sqrt{1 + g_5/q^2}$, that is given in Ref. \cite{Biancalani17}, gives the same results. To describe how the frequency changes with the radial wavenumber, a  polynomial has been taken. 
Moreover, to take into account the frequency saturation for higher wavenumbers\cite{Singh17}  we have introduced a function of the form $1 / (1 + g_4 k)$. Here, $k = k_r \rho_{i}$, $v_{Ti} = \sqrt{q_eT_i/2m_p}$. The resulting frequency interpolating formula is the following one:
\aeq
f_{\omega} \left[\yfrac{\sqrt{2} v_{Ti}}{R} \right] = \yfrac{g_1 + g_2 k^2 + g_3 k^4}{1 + g_4 k} \yfrac{\exp\left( -g_5 q^2 \right)}{1 + g_6 e}.
\label{formula:FIT_w}
\eeq
Among different tested functions, this form gives the best approximation to numerically simulated values of the GAM frequency, it has one of the smallest 95\% confidential bounds and is not overfitted.
The corresponding coefficients $g$ with their 95\% confidential bounds (lower $g_{lc}$ and upper $g_{uc}$ bounds) are
\aeqns
g =      &&[3.7733,\ 6.3505,\ -1.9741e1,\ 1.3557e-1,\ 1.4620e-3,\ 1.1684],\\*
g_{lc} = &&[3.6745,\ 3.3168,\ -2.8800e1,\ -6.0078e-2,\ 1.1373e-3,\ 1.1234],\\*
g_{uc} = &&[3.8720,\ 9.3843,\ -1.0682e1,\ 3.3121e-1,\ 1.7866e-3,\ 1.2135].
\eeqns
Results for the Eq. (\ref{formula:FIT_w}) are depicted in Fig. \ref{fig:FIT_w}.

For the damping rate we have derived the following expression (here, the damping rate is normalized to $\sqrt{2} v_{Ti}/R$):
\aeq
f_{\gamma} \left[\yfrac{\sqrt{2} v_{Ti}}{R} \right] = \yfrac{\left(h_1 + h_2 k^2\right)\exp\left[-h_3 q^2\right]}{1 + h_4 e^2} + \yfrac{\left(h_5 + h_6 k^2\right)\exp\left[-h_7 q^2 \right]}{1+h_8 e^4}.
\label{formula:FIT_g}
\eeq
with interpolating coefficients
\aeqns
h =&& [-1.2494e-2,\ -8.9688e-1,\ 4.5498e-2,\ -1.9884e-1,\\* 
&&-1.1248e-2, -2.5481,\ -5.3340e-3,\ 7.7748e-1],\\
h_{lc} =&& [-2.3115e-2,\ -1.6490,\ 2.5215e-2,\ -3.3573e-1,\\* 
&&-2.5523e-2, -3.1909,\ -1.9665e-2,\ 5.1924e-2],\\
h_{uc} =&& [-1.8723e-3,\ -1.4471e-1,\ 6.5781e-2,\ -6.1955e-2,\\* 
&&3.0272e-3, -1.9053,\ 8.9973e-3,\ 1.5030].
\eeqns
Comparison between the results from the gyrokinetic simulations and the interpolation expression for the GAM damping rate is  shown in Fig. \ref{fig:FIT_g} for some specific values of parameters taken as examples.
\section{Phase mixing}
\label{PhaseMixing}
\begin{figure}[t!]
\centering
\includegraphics[scale = 0.15]{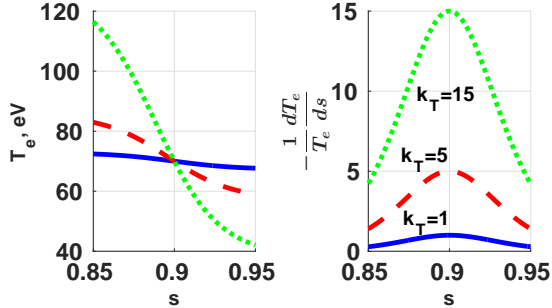}
\caption{\label{fig:PhM_TemperProfs}Temperature and temperature gradient radial profiles for different $k_T$: $k_T = [1, 5, 15]$.}
\end{figure}

\begin{figure}[t!]
\centering
\includegraphics[scale = 0.15]{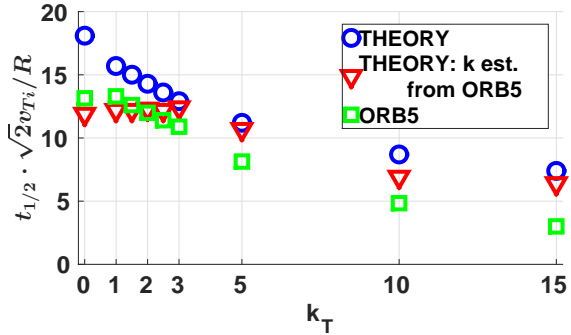}
\caption{\label{fig:PhM_ScanKappa} Dependence of the GAM half-decay time on the temperature gradient obtained from the simulations in ORB5 (green squares), from the theory using a linear estimation Eq. (\ref{equ:K_evol}) (blue dots) and estimation from ORB5 (red triangles) of the radial wavenumber.}
\end{figure}

To investigate the influence of the phase mixing on the GAM dynamics the same parameters as described in chapter \ref{PARS_WOMIX} has been used, but the temperature gradient (the same for both the electrons and ions to have $\tau_e = T_e/T_i = 1$, $T_i(s_0) = 70$ eV) at a radial position $s_0 = 0.90$ has been introduced. The radial point $s_0 = 0.90$ has been chosen here to be in agreement with the section \ref{PARS_WOMIX}. Initial radial wavenumber of the radial electric field is $k = 0.108$. The safety factor is $q(s_0) = 4.0$.
We consider a temperature profile of the following form, similarly to Ref. \cite{Biancalani16}:
\aeq
\yfrac{T_e(s)}{T_e(s_0)} = \exp\left[-\Delta \cdot k_T \cdot \tanh\left(\yfrac{s - s_0}{\Delta}\right)\right],
\eeq
where $\Delta = 0.04$, $k_T = - \left. d[\ln(T)]/ds \right|_{s = s_0}$. The temperature profiles and the corresponding temperature gradient profiles for different $k_T$ in a radial interval $s = [0.85, 0.95]$, are shown in Fig. \ref{fig:PhM_TemperProfs}. Dependence of the GAM half-decay time $t_{1/2}$ on the temperature gradient has been investigated in the domain $k_T \in [1, 15]$.
A scan of gyrokinetic simulations with the temperature gradient $k_T$ has been performed, and the results are depicted in Fig. \ref{fig:PhM_ScanKappa}.
In presence of a temperature gradient, the GAM is observed to oscillate with different frequencies at different radial points, that leads to the distortion of the initial GAM radial structure. Producing higher radial wavenumbers, this distortion amplifies the GAM damping.
This combined effect, already investigated for a more simplified configuration in Ref. \cite{Palermo16, Biancalani16}, has been observed even more pronounced in the simulations described here.
In fact, here the phase mixing effect is investigated using gyrokinetic simulations with kinetic electrons that significantly influences the GAM damping, and, as a consequence, the GAM half-decay time. 
For example, using the Sugama-Watanabe model\cite{Sugama08}, which is derived with adiabatic electrons, for the Landau damping and combining with phase mixing, we have obtained $t_{1/2}[R/ \sqrt{2} v_{Ti}] = 118$ for the $k_T = 1$ and $t_{1/2}[R/ \sqrt{2} v_{Ti}] = 23.4$ for $k_T = 10$, that predicts much longer half-decay time of the GAM in comparison to the calculations based on the simulations with the kinetic electrons (compare with a Fig. \ref{fig:PhM_ScanKappa}).

In order to verify the results of gyrokinetic simulations, we have used a  theoretical simplified model of the phase mixing, proposed in Ref.~\cite{Zonca08, Palermo16, Biancalani16}, where the linear growth in time of the radial wavenumber is considered. 

In the phase mixing simulations a space point $s_0$ is considered with a certain temperature $T(s_0)$ and temperature gradient $k_T(s_0)$. 
Initial radial electric field has the following radial structure:
\aeq
E(s) = E_0 \cos(k_0 s)
\eeq
with an initial amplitude $E_0$ and initial normalized radial wavenumber $k_0$.
The electric field is assumed to evolve in time at a point $s_0$ according to a simple rule  
\aeq
E(s_0, t)  = E_{a}(s_0, t) \cos(\omega(s_0) t), 
\eeq
where $E_{a}(s_0, t)$ is an amplitude of  the electric field, that changes in time due to the damping, $E_{a}(0) = E_0$.
The general form of the GAM frequency is
\aeq
\omega(s, t) = \sqrt{\frac{2 T_e(s)}{2m_p}} \omega^*(k, q, e), 
\eeq
where $\omega^*(k, q, e)$ describes frequency dependence on the radial wavenumber, the safety factor and the elongation. The safety factor profile is taken to be flat, and plasma with a circular cross-section is considered: $e = 1.00$, $r \approx a s$.

The damping rate is defined as
\aeq
\gamma(s_0, t) = \yfrac{1}{E(s_0,t)}\yd{E(s_0,t)}{t}.
\eeq

At the beginning of every time interval $[t_1, t_1 + \Delta t]$, new values of the damping rate $\gamma(s_0, t_1)$ and frequency $\omega(s_0, t_1)$ are found with the scaling formulae given in Eq. (\ref{formula:FIT_g}) and (\ref{formula:FIT_w}), using a current value of the wavenumber $k(s_0, t_1)$. 
A new value of the electric field can be found, assuming that the damping rate is constant at the lapse of time $[t_1, t_1 + \Delta t]$:
\aeq
E(s_0, t_1 + \Delta t) = E(s_0, t_1)\cdot(1 + \gamma(s_0, t_1) \Delta t).
\eeq

After that, a new value of the wavenumber $k(s_0, t_1+\Delta t)$ is calculated using the radial derivative of the frequency 
\aeq
\left. \fpar{\omega(s, t_1)}{s} \right|_{s = s_0} = - \yfrac{1}{2} \omega(s_0, t_1) k_T.
\eeq
With that, the wavenumber is assumed to change linearly in time as
\aeq
k(s_0, t_1 + \Delta t) = k(s_0, t_1) - \sqrt{2} \rho^* \left. \fpar{\omega(s, t_1)}{s} \right|_{s = s_0} \Delta t,
\label{equ:K_evol}
\eeq
where $\rho^* = \rho_s / a$, $\rho_s = c_s/ \omega_{ci}$.
Another option, it is to estimate the time evolution of the radial wavenumber directly from numerical calculations in ORB5. Substituting new value of the normalized wavenumber $k(s_0, t_1+\Delta t)$ into Eq. (\ref{formula:FIT_g}),  we can find the damping rate $\gamma(s_0, t_1+\Delta t)$ at the next time point.

The results obtained with this reduced theoretical model are also shown in Fig. \ref{fig:PhM_ScanKappa}. As it can be seen in Fig. \ref{fig:PhM_ScanKappa}, the 
qualitative dependence of the half-decay time on the temperature 
gradient finds a good match of gyrokinetic simulations of ORB5 and analytical 
theory. The difference is due to the global dynamics of the ORB5 
simulations, which is compared here with a theory where the phase mixing 
follows a local estimation given in Ref.~\cite{Zonca08}.
\section{Comparison with experimental data}
The dispersion relations obtained in Sec. \ref{ch:InterpExp} as 
an interpolation of gyrokinetic simulations and given in Eqs. (\ref{formula:FIT_w}), (\ref{formula:FIT_g}) can be used to compare numerical estimations of the GAM behaviour to measurements of the GAM frequency, performed on ASDEX Upgrade tokamak \cite{Conway08} using Doppler reflectometry. 
More precisely, we consider the discharge AUG\#20787 with the plasma elongation at the edge $e = 1.09$ (and we assume that it is constant at the considered radial region $\rho = r/a = [0.8, 1.0]$). 

\begin{figure}[t!]
\begin{subfigure}[t]
{0.49\textwidth} 
\includegraphics[width=\textwidth]{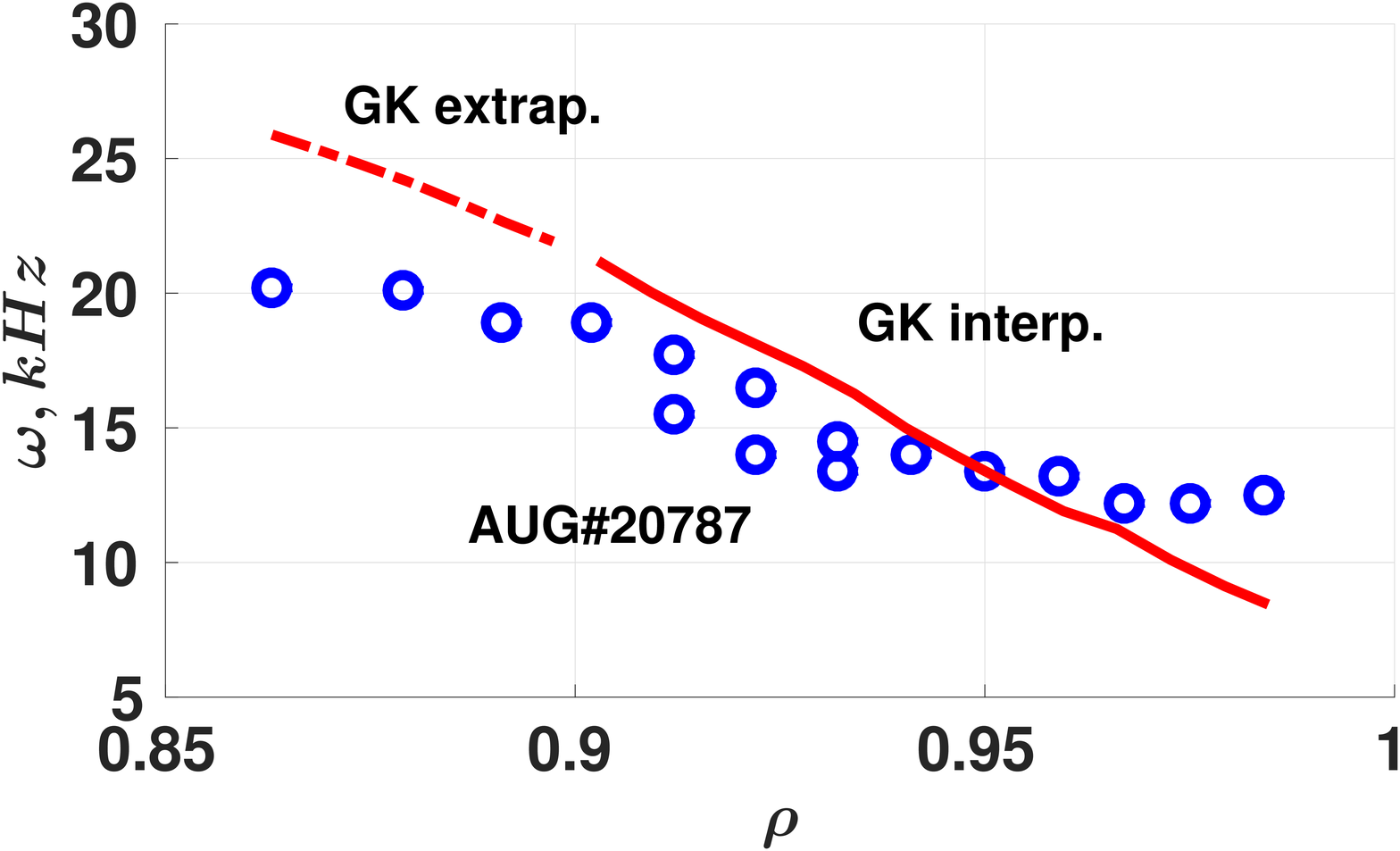}
\end{subfigure}
\begin{subfigure}[t] 
{0.49\textwidth} 
\includegraphics[width=\textwidth]{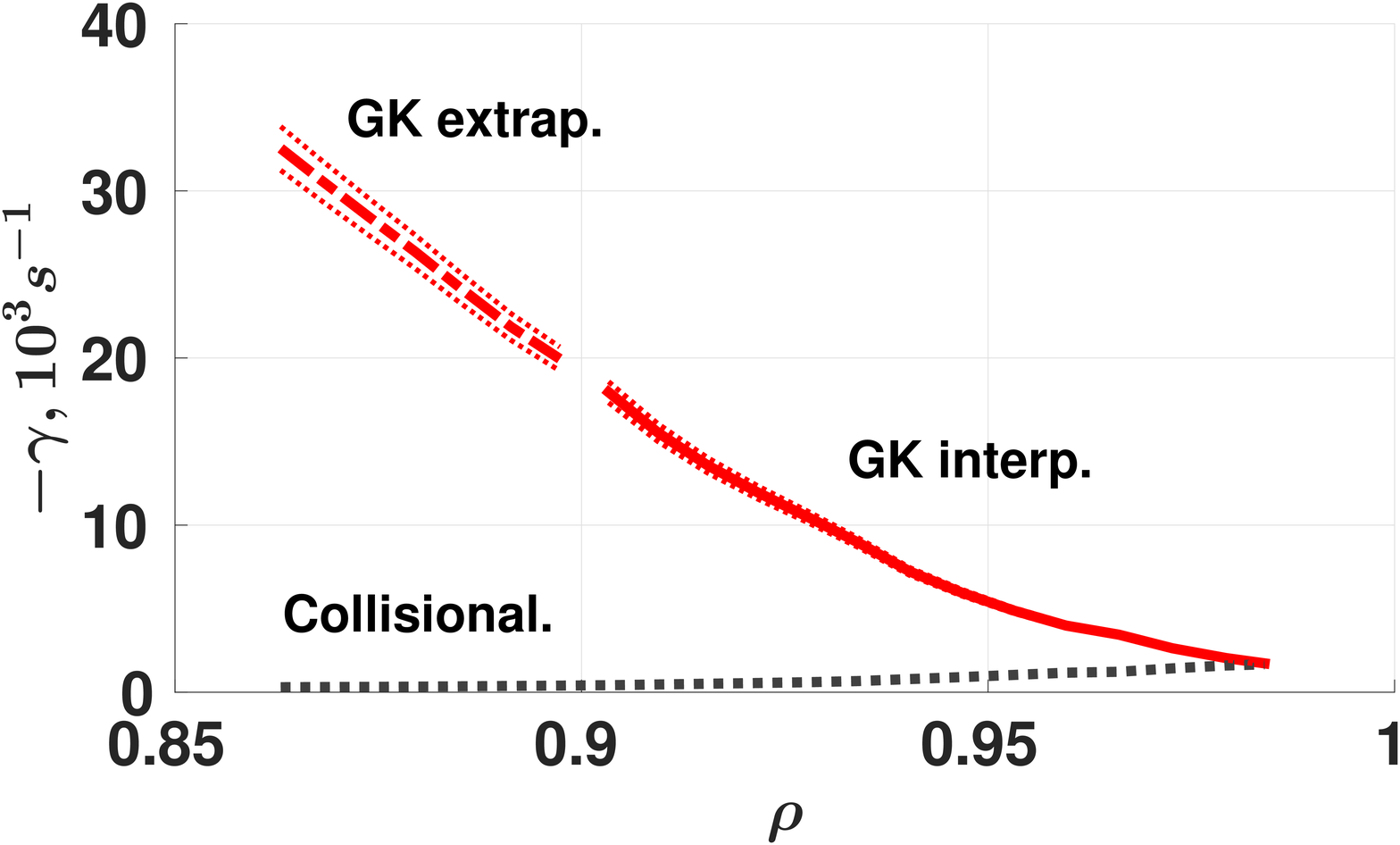}
\end{subfigure}
\caption{\label{fig:PhM_Comp_woMix_shot20787} Comparison of the experimental GAM frequencies \cite{Conway08} to the numerical values, obtained with the formula given in Eq. (\ref{formula:FIT_w}). Numerical damping rate is depicted on the right plot. The grey dotted line is an estimation of the collisional damping rate of the GAM found using expressions given by Gao in Ref. \cite{Gao2013}. The red dotted lines are the 95\% confidential bounds of the approximated damping rate.}
\end{figure}

The GAM radial wavenumber is considered to be constant and is estimated to be $k_ra = 40\pi$ from the experimental radial profile of the GAM amplitude (see Fig. 5f in Ref. \cite{Conway08}). Experimental safety factor and ion temperature profiles have been taken to estimate the GAM frequency and damping rate using the scaling formulae (\ref{formula:FIT_w}), (\ref{formula:FIT_g}) at different radial points $\rho$.
In Fig. \ref{fig:PhM_Comp_woMix_shot20787} the GAM frequency profiles with corresponding theoretical prediction are depicted. A good general agreement is found in the central region of interest, where the GAM intensity, measured in the experiments, is peaked. On the other hand, the linear dispersion relation \ref{formula:FIT_w} can not explain neither the staircase nature (the plateaus) of the frequencies nor the GAM peak splitting that is observed experimentally at the radius positions $\rho = 0.922$ or $\rho = 0.932$ (although the presence of GAM eigenmodes has been suggested by simplified analytical 
models \cite{Sasaki08, Ilgisonis14}, whose detailed analysis is out of 
the scope of this paper).
For this reason, we can conjecture that the coherent phenomena at the basis of the formation of GAM extended eigenmode or frequency splitting must have a nonlinear origin.

For reference we have given here estimation of the GAM collisional damping rate using formulae, derived by Gao in Ref. \cite{Gao2013}. Introducing normalized ion collision rate $\hat{\nu}_i = \nu_i  qR / v_{ti}$, the collisional damping rate is calculated as\cite{Gao2013}
\aeq
\yfrac{\gamma^{col}}{v_{Ti}/qR} = - \yfrac{3\hat{\nu}_i}{14 + 8\tau_i},
\eeq
if $\hat{\nu}_i \ll 1$, and as:
\aeq
\yfrac{\gamma^{col}}{v_{Ti}/qR} = - \yfrac{3}{8} \hat{\nu}_i \left(\yfrac{7}{4} + \tau_i + \yfrac{\hat{\nu}_i^2}{q^2} \right)^{-1},
\eeq
if $\hat{\nu}_i \geq 1$. To find the ion collisional rate we have used classical expressions:
\aeqn
\nu_i &=& 4.8\cdot10^{-8} Z^4 \mu^{-1/2} n_i[cm^{-3}] T_i[eV]^{-3/2}\ln\Lambda,\\
\ln\Lambda &=& 23 - \ln \left[ \sqrt{2n_i[cm^{-3}]}\yfrac{Z^3}{T_i[eV]^{3/2}} \right],
\eeqn
where $\mu \equiv m_i / m_p = 2$, $Z = 1$.

According to the Fig. \ref{fig:PhM_Comp_woMix_shot20787}, the collisional damping is found to be negligible in the radial domain where GAMs are experimentally 
measured, except in a very narrow region close to the separatrix, where 
it can be of the same order of magnitude as the Landau damping.
\section{Conclusions}
In tokamak plasmas, the drift-wave turbulence gives rise to the zonal flows that in their turn shear and distort convective and turbulent cells leading to the saturation of turbulence and, consequently, to a reduction of the radial heat transport. Action of the magnetic curvature results in the oscillatory zonal flows, so-called geodesic acoustic modes. The peculiarity of the GAM oscillations resides in the different shearing efficiency that the ZF have in relation to their oscillatory behavior.
The nonlinear interactions between the GAM and the DW turbulence is defined in a high degree by the GAM damping rate. Lack of the experimental data of this characteristic of the GAM makes the results from linear gyrokinetic simulations particularly important for analytical and numerical investigation of the nonlinear GAM-DW systems.

In this work, linear gyrokinetic simulations have been performed with kinetic electrons to study the GAM dynamics. Numerical results have been compared to analytical theories, derived with adiabatic electrons. It has been shown that analytical theories, derived with adiabatic electrons, result in smaller values of the  damping rate and for higher wavenumbers diverge from numerical calculations of the frequency.  
That is why, investigating the GAM dependence on the plasma safety factor, elongation and radial wavenumber, we have found approximating analytic expressions for the frequency and damping rate to predict the GAM behaviour in different plasma regimes. 
The derived expressions can be used to estimate the GAM linear characteristics used in analytical models of the nonlinear interactions between the GAM and the DW, such as different reduced models \cite{Chen00, Zonca08}.
Using these formulae, the phase mixing effect on the damping rate has also been calculated. Based on the gyrokinetic simulations with kinetic electrons, the results have shown smaller half-decay times of the GAM in comparison with the Ref. ~\cite{Palermo16,Biancalani16}. 

The GAM is one of the special features of the I-mode and can be observed in the L-mode \cite{Conway11, Manz2015, Wang13}.
Comparison of the characteristic drive time of the GAM $t_{RD} \sim 1/\gamma_{RD}$, which is given by the nonlinear coupling with the ion-temperature-mode (ITG)\cite{Zonca08, Chen00}, with the GAM half-decay time $t_{1/2}$ confirms the results of the Ref. ~\cite{Palermo16,Biancalani16}. 
Indeed, we estimate the GAM drive time to be $t_{RD} < t_s$ (where $t_s \sim 2^{-1/2} R/v_{Ti}$) in the L-mode, $t_{RD} \sim t_s$ in the I-mode and $t_{RD} \sim 10t_s$ in the H-mode, according to Ref. ~\cite{Biancalani16}. In this case, it can be seen from the Fig. \ref{fig:PhM_ScanKappa} that the GAM half-decay time, which is defined by both the Landau damping and the phase mixing effect, is much higher than the drive-time in the L and I modes, $t_{1/2} > t^{L,I}_{RD}$, for all considered values of $k_T$. This means that the energy transfer rate from the ITG turbulence to the GAM exceeds the Landau-phase mixing damping rate of the GAM. 
As a result, the GAM can be observed in the L and I modes, but not in the H-mode, where  $t_{1/2} < t^H_{RD}$ already for $k_T > 3$. 
This could be the explanation for the result that the GAM are not observed in the high-confinement mode, as proposed in Ref. \cite{Palermo16, Biancalani16}.

The approximating expressions have showed quite good agreement with experimental data, but estimations of the GAM frequency, obtained from linear gyrokinetic simulations, does not explain the staircase radial profile of the frequency and the GAM peak splitting. The frequency expression describes only the continuum or dispersive mode in contrast to eigenmode. The latter is characterized by the GAM mode frequencies which are predicted to remain constant over a large radial extent, but a significant radial overlap in the frequency radial profile can be observed, that lead to the GAM frequency peak splitting\cite{Wang13}.
\section*{Acknowledgments}
Part of this work was done while two of the authors, I. Novikau and A. Biancalani, were visiting LPP-Palaiseau (France), whose team is kindly acknowledged for the hospitality. 
The authors acknowledge discussions with F. Jenko, L. Villard, X. Garbet and T. G\"orler.

\appendix
\section{Numerical convergence tests}
\label{appendix:convergence}

\begin{figure}[b!]
\centering
\includegraphics[scale = 0.15]{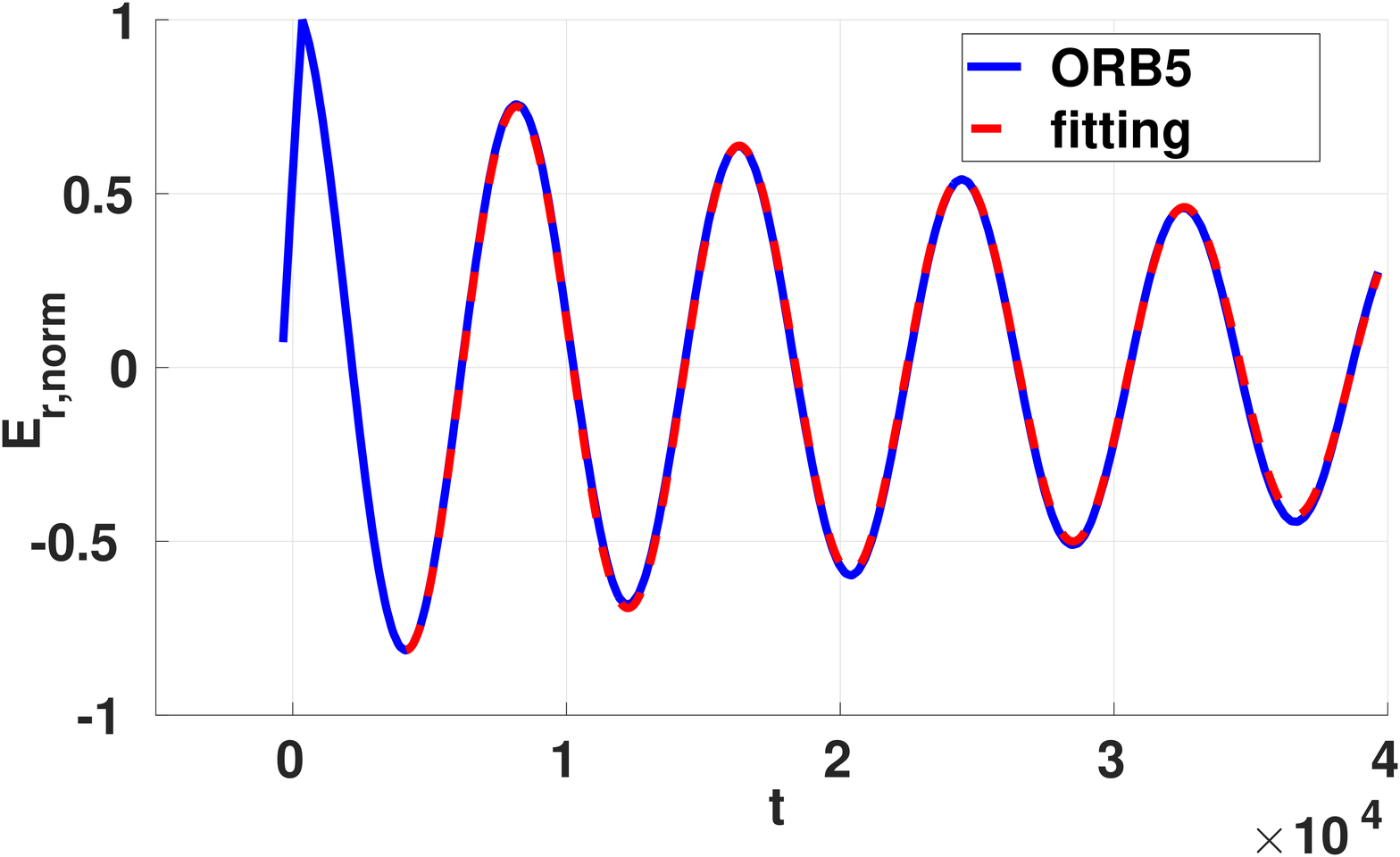}
\caption{\label{fig:Show_fitting} Fitting of the normalized radial electric field for the case: $e = 1.30$, $q = 4.0$, $k = 0.108$. Here, the blue line is a filtered (from high-frequency Alfv\'en oscillations) smoothed signal of poloidally averaged radial electric field from ORB5. The red one is the fitting. Here, the time is normalized to $\omega_{ci}^{-1}$.}
\end{figure}

\begin{figure}[t!]
\begin{subfigure}
{0.49\textwidth} 
\includegraphics[width=\textwidth]{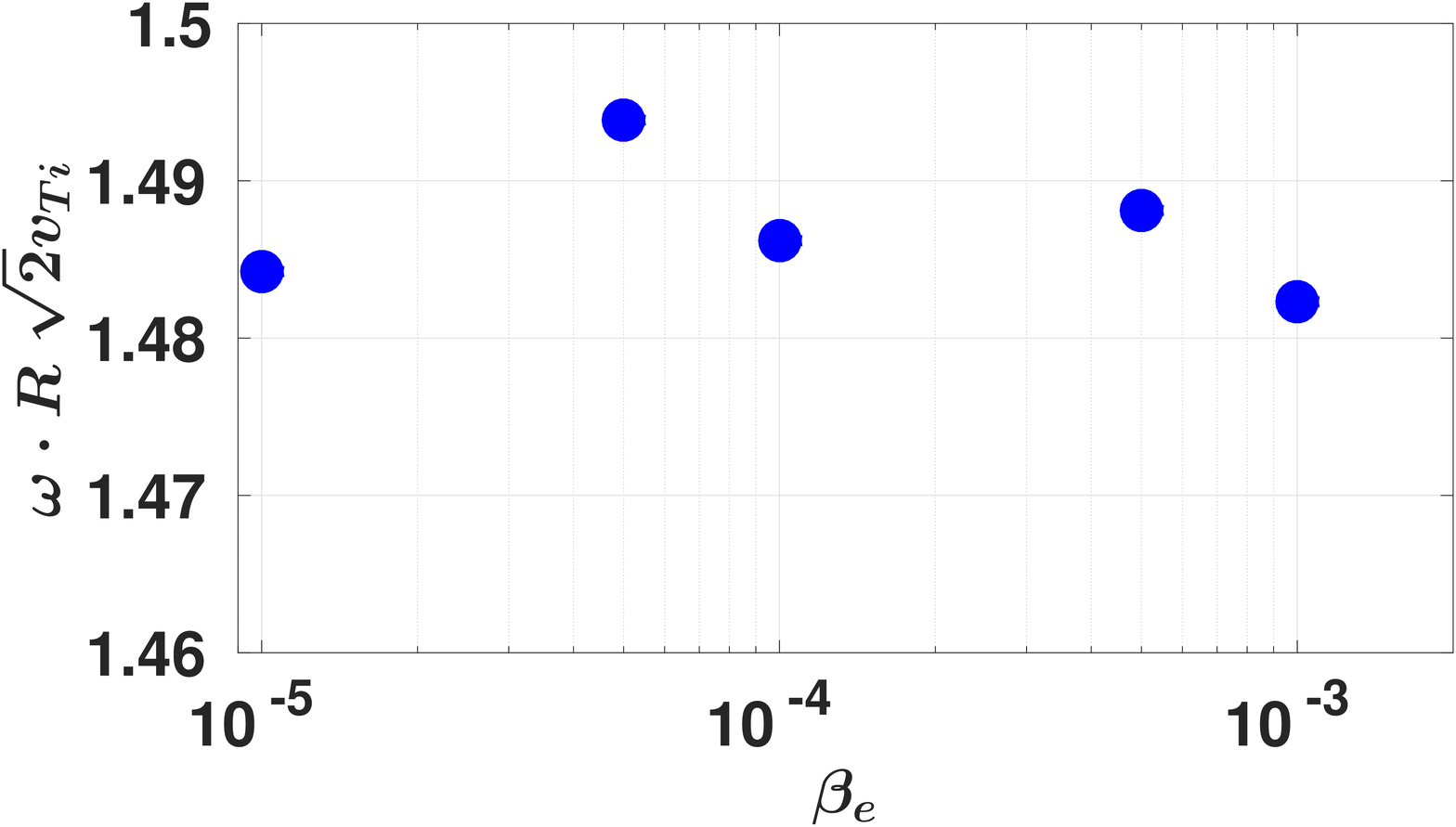}
\end{subfigure}
\begin{subfigure}
{0.49\textwidth} 
\includegraphics[width=\textwidth]{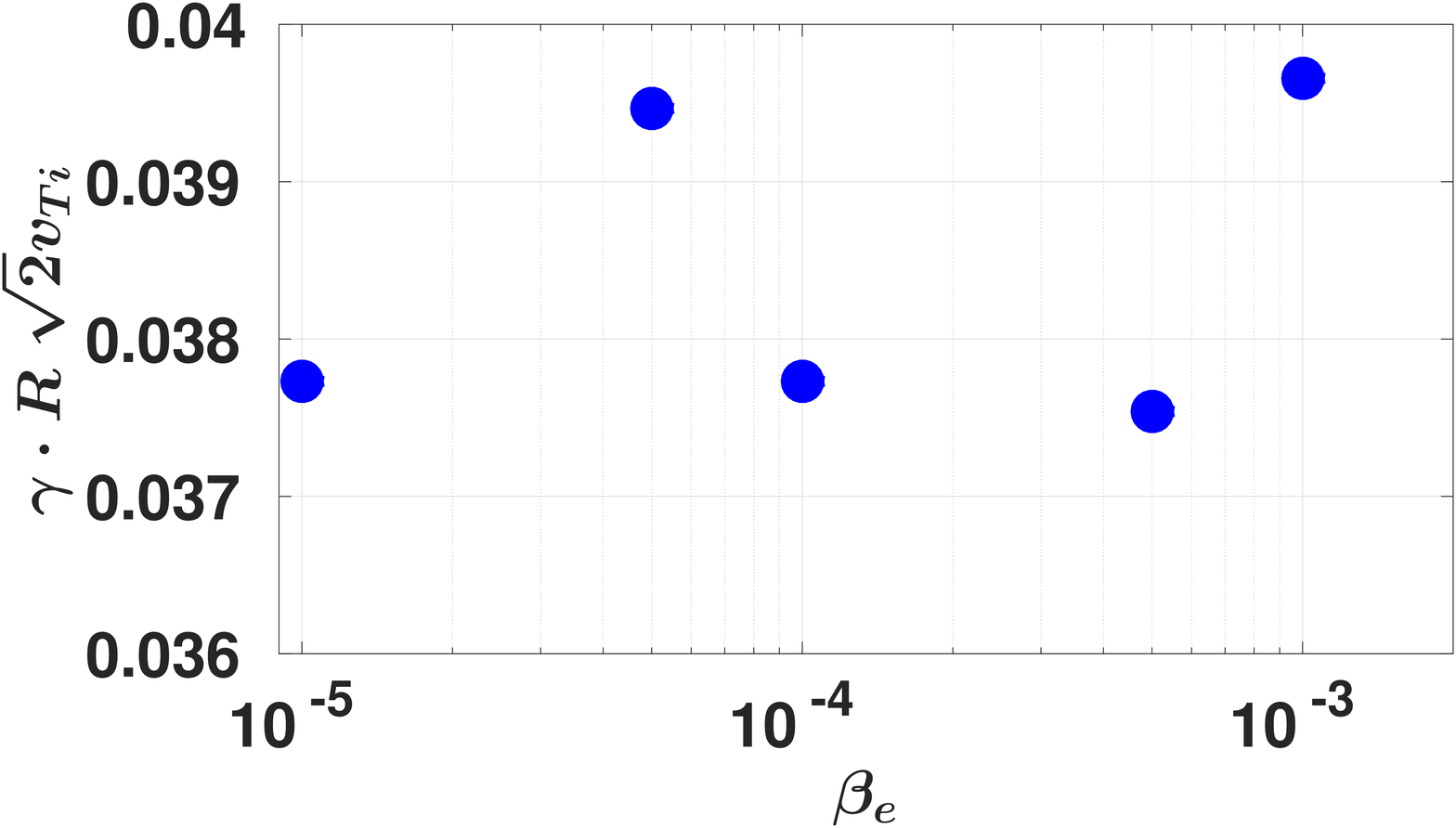}
\end{subfigure}
\caption{\label{fig:Beta_scan} Scan of the GAM frequency and damping rate on different values of the electron beta in linear electromagnetic collisionless simulations with kinetic electrons. Here, $T_e = 70$ eV, $B = 2$ T, $k = 0.108$, $q = 4.0$, $e = 1.30$.}
\end{figure}
To calculate the GAM damping rate and frequency, poloidally averaged radial electric field has been fitted, using the Levenberg\textendash Marquardt algorithm \cite{NumRec02}, to a function of the form $\exp(\gamma t) \cos(\omega t)$, where $\gamma$, $\omega$ are sought-for damping rate and frequency. Before the fitting it's necessary to filter the radial electric field to get ride of the high-frequency Alfv\'en oscillations. In Fig. \ref{fig:Show_fitting} the fitting is depicted for the case: $e = 1.30$, $q = 4.0$, $k = 0.108$. It is worth to mention that the choice of a time interval, where the fitting is performed, can influence the result damping rate, and it is not so crucial for the GAM frequency calculation. This ambiguity in the choice of the time interval can be explained by the fact that at the beginning of the simulations there are some transient processes that must be excluded from the damping rate measurements. Moreover, with the time, the global effects start to play a significant role, distorting initial radial structure of the radial electric field, that makes the damping rate to be variable in time.

\begin{figure}
\centering
\includegraphics[scale = 0.18]{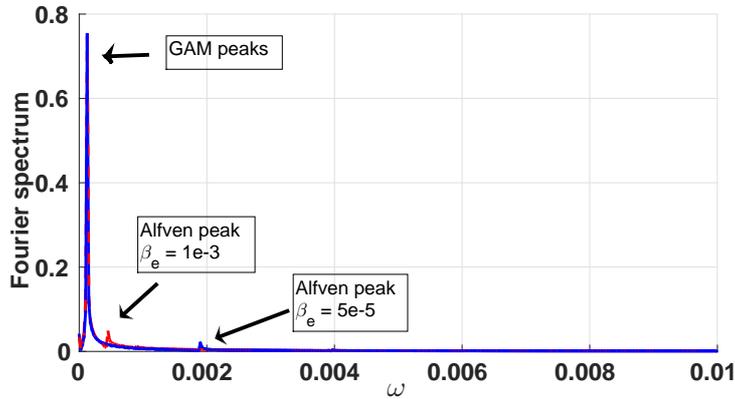}
\caption{\label{fig:Beta_FS}Fourier spectrum of the radial electric field for different values of the electron beta: $\beta_e = 5e-5$ (blue line), $\beta_e = 1e-3$ (red line). Only the Alfv\'en frequency changes with the electron beta. The GAM frequency remains the same. Here, the frequency is normalized to $\omega_{ci}$.}
\end{figure}

The GAM dynamics (frequency and damping rate) doesn't depend on the plasma density in linear calculations (see Fig. \ref{fig:Beta_scan} and \ref{fig:Beta_FS}). But the frequency  of the Alfv\'en waves decreases with the increase of the density, and for high values of the plasma density it becomes difficult to separate acoustic and Alfv\'enic time scales (see Fig. \ref{fig:Beta_FS}).

\begin{figure}[t!]
\begin{subfigure}[t] 
{0.49\textwidth} 
\includegraphics[width=\textwidth]{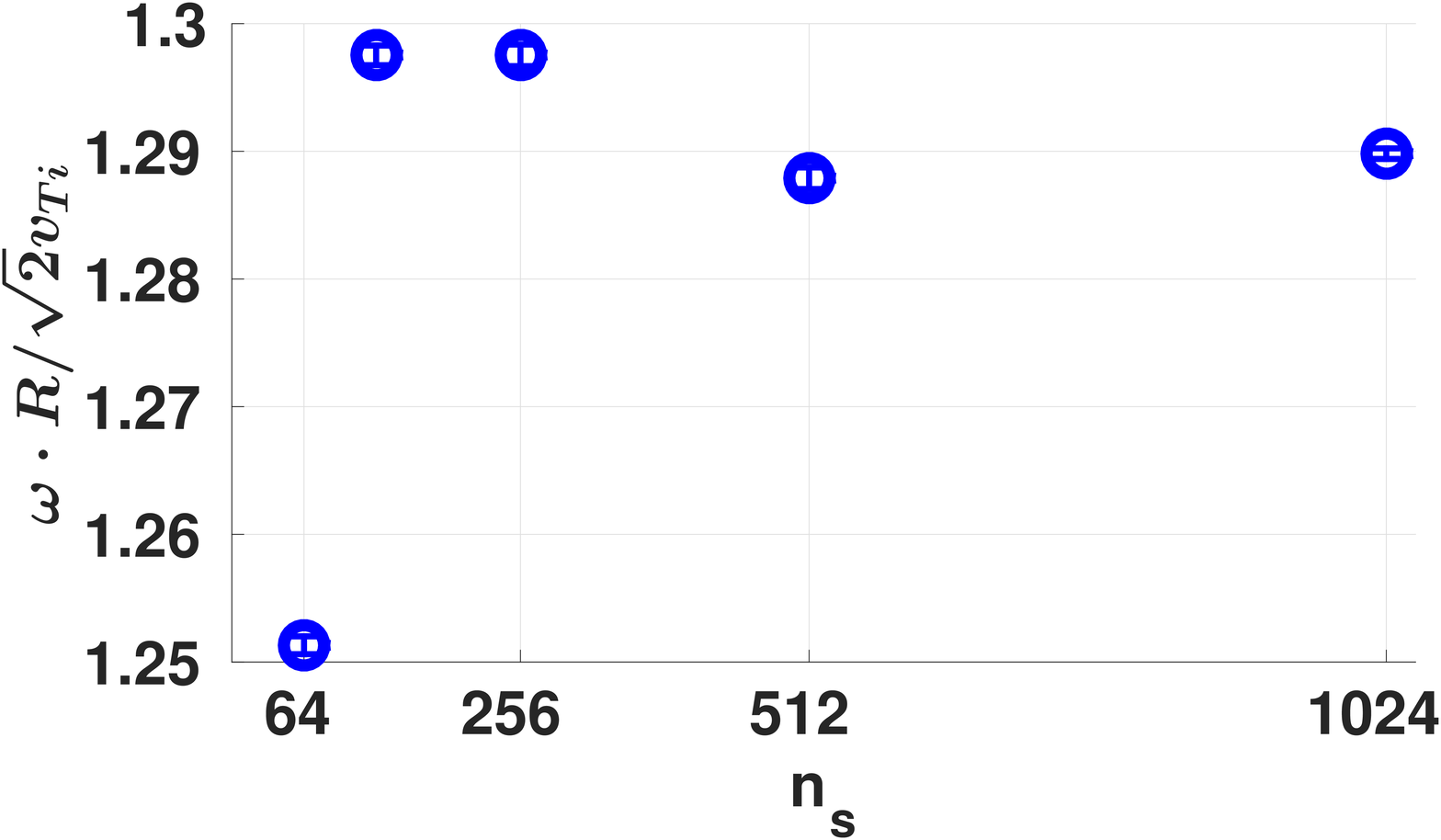}
\caption{\label{fig:ConvTests_ns_w}}
\end{subfigure}
\begin{subfigure}[t]
{0.49\textwidth} 
\includegraphics[width=\textwidth]{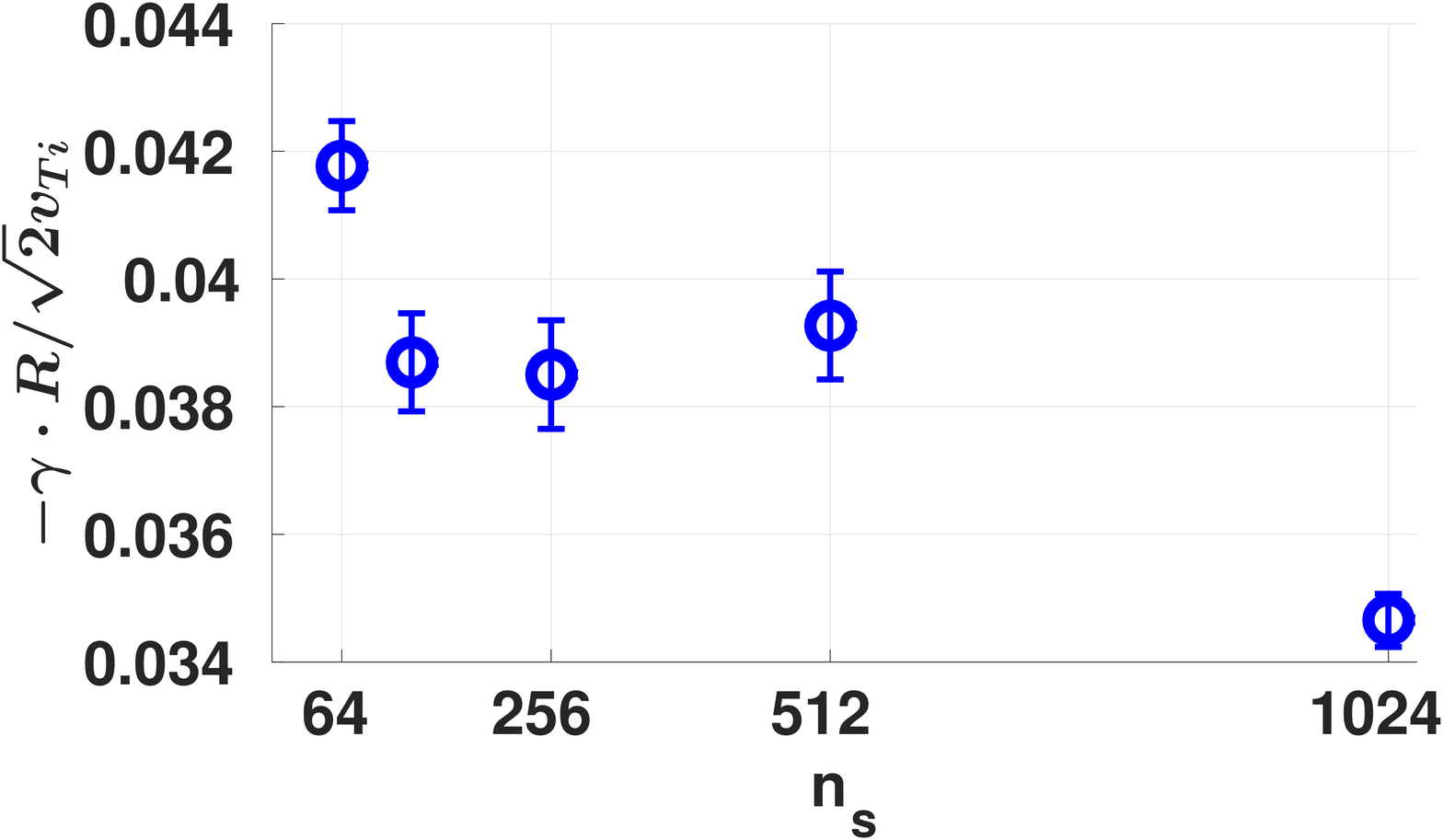}
\caption{\label{fig:ConvTests_ns_g}}
\end{subfigure} 
\begin{subfigure}[t]  
{0.49\textwidth} 
\includegraphics[width=\textwidth]{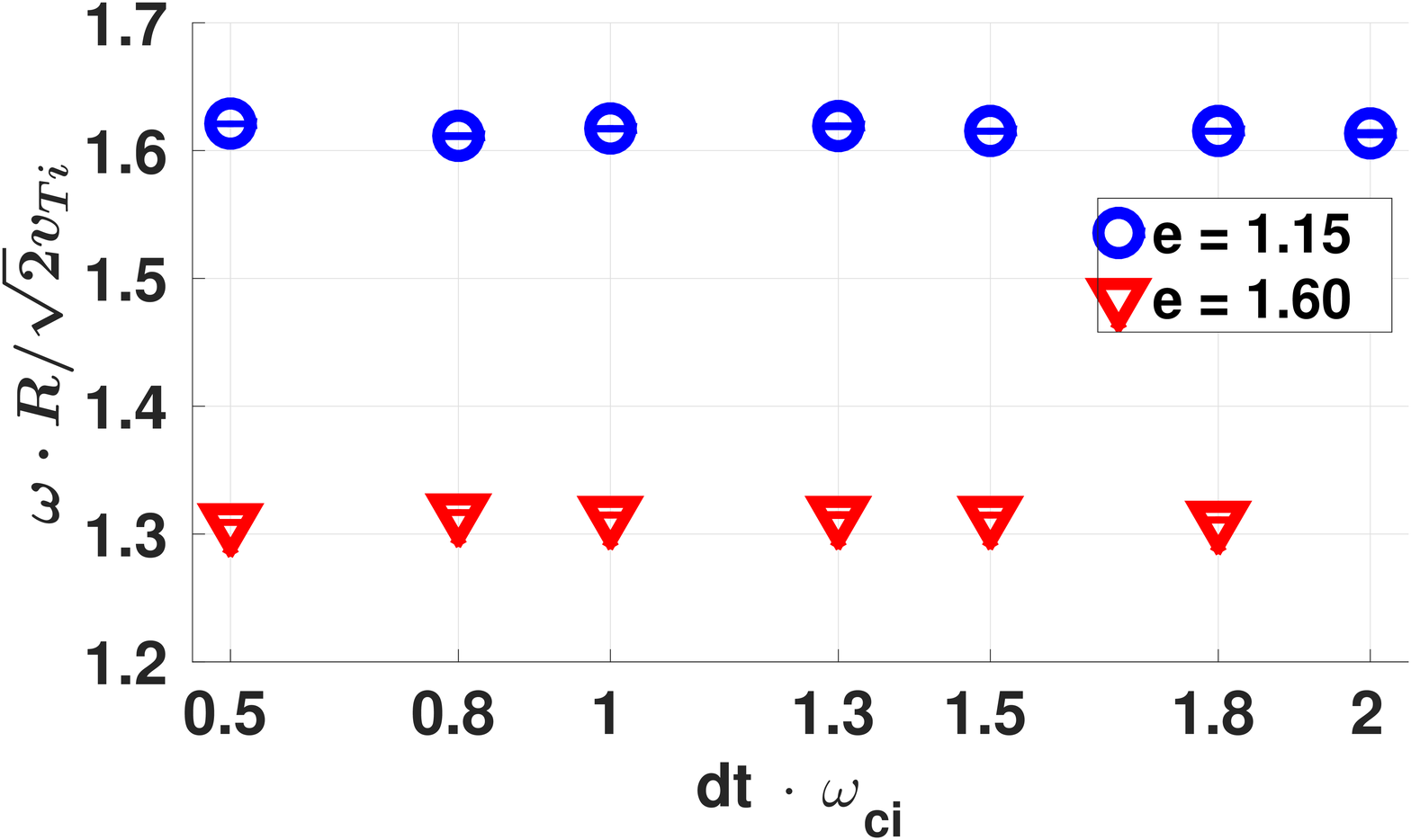}
\caption{\label{fig:ConvTests_dt_w}}
\end{subfigure}
\begin{subfigure}[t]
{0.49\textwidth} 
\includegraphics[width=\textwidth]{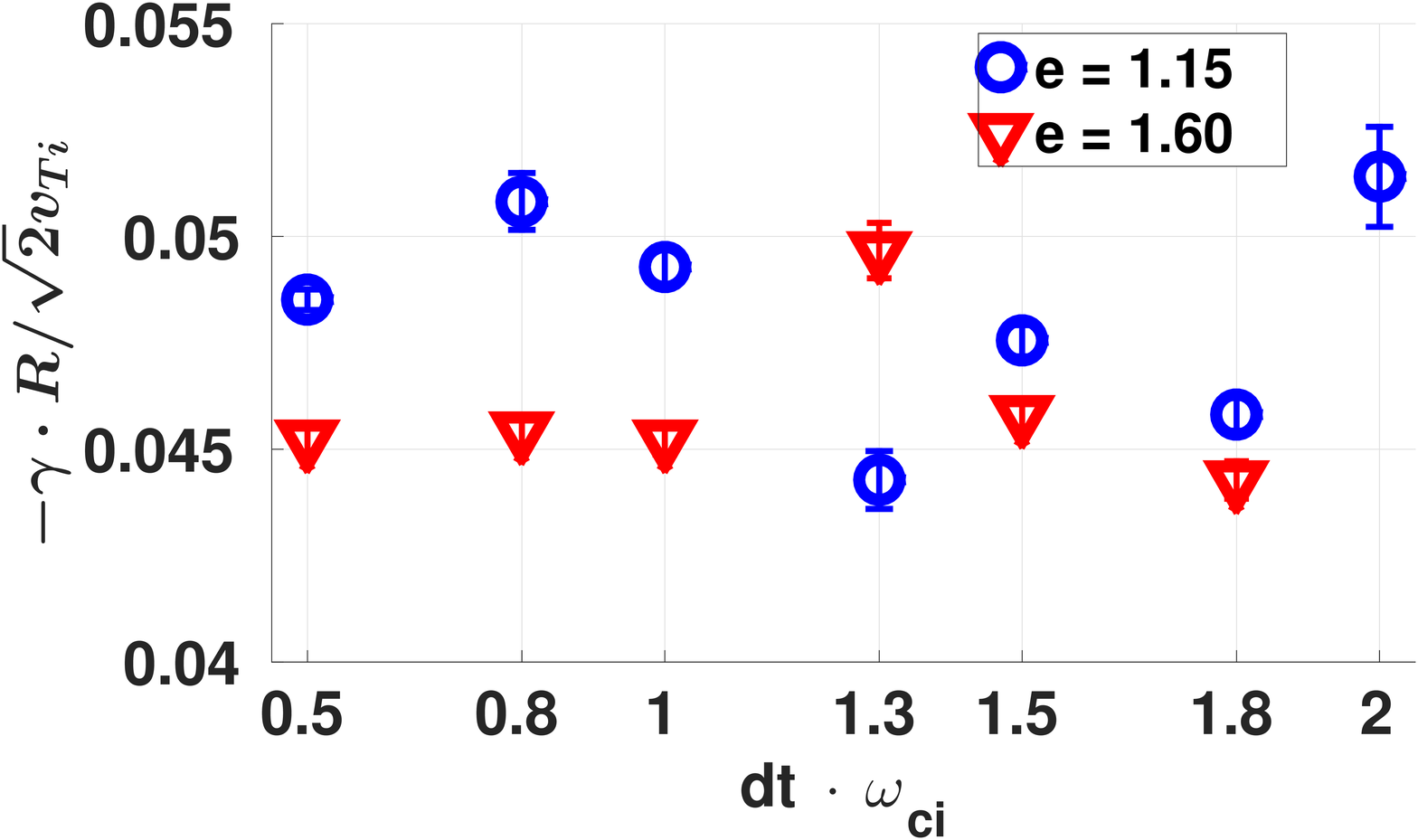}
\caption{\label{fig:ConvTests_dt_g}}
\end{subfigure}
\begin{subfigure}[t]  
{0.49\textwidth} 
\includegraphics[width=\textwidth]{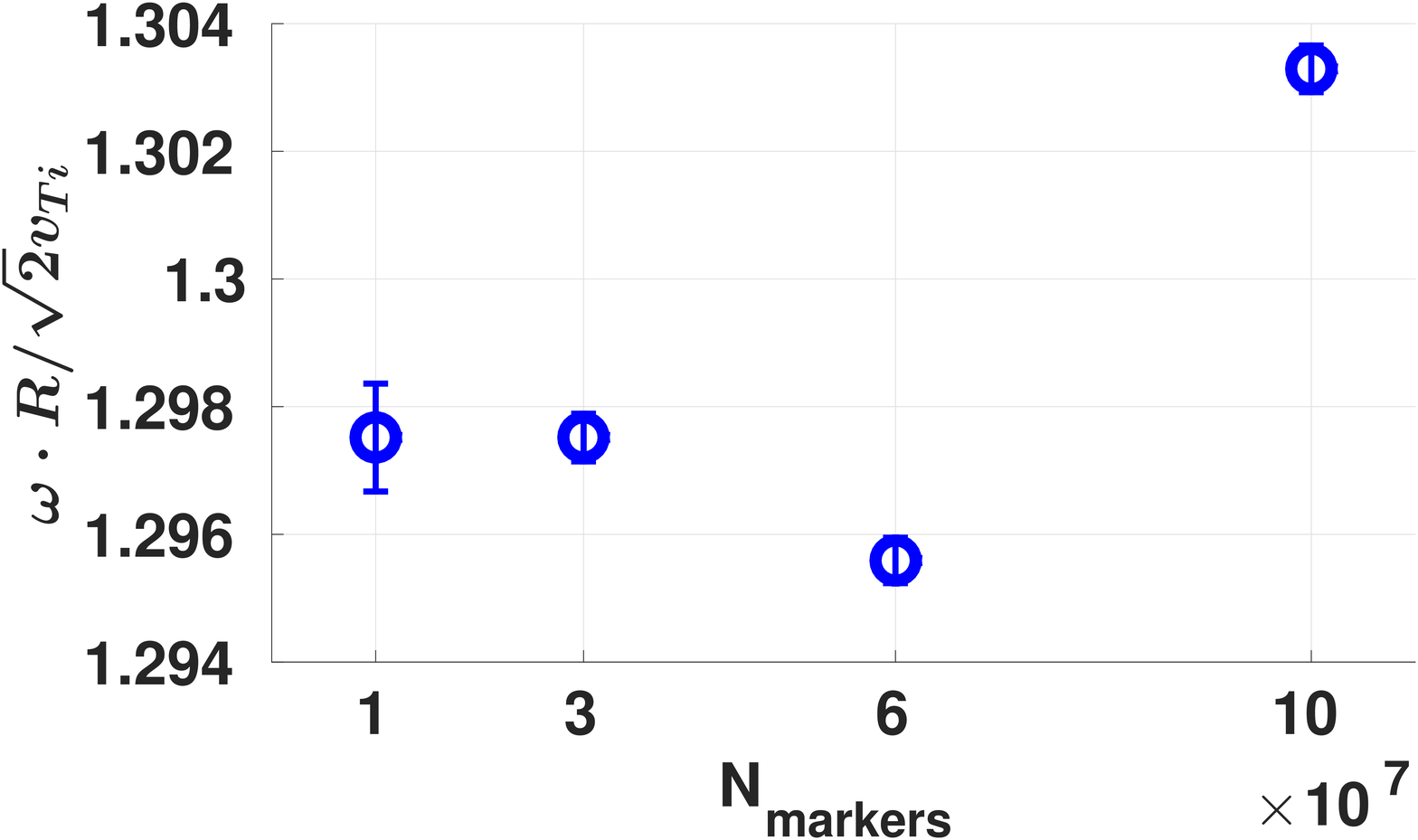}
\caption{\label{fig:ConvTests_nptot_w}}
\end{subfigure}
\begin{subfigure}[t]
{0.49\textwidth} 
\includegraphics[width=\textwidth]{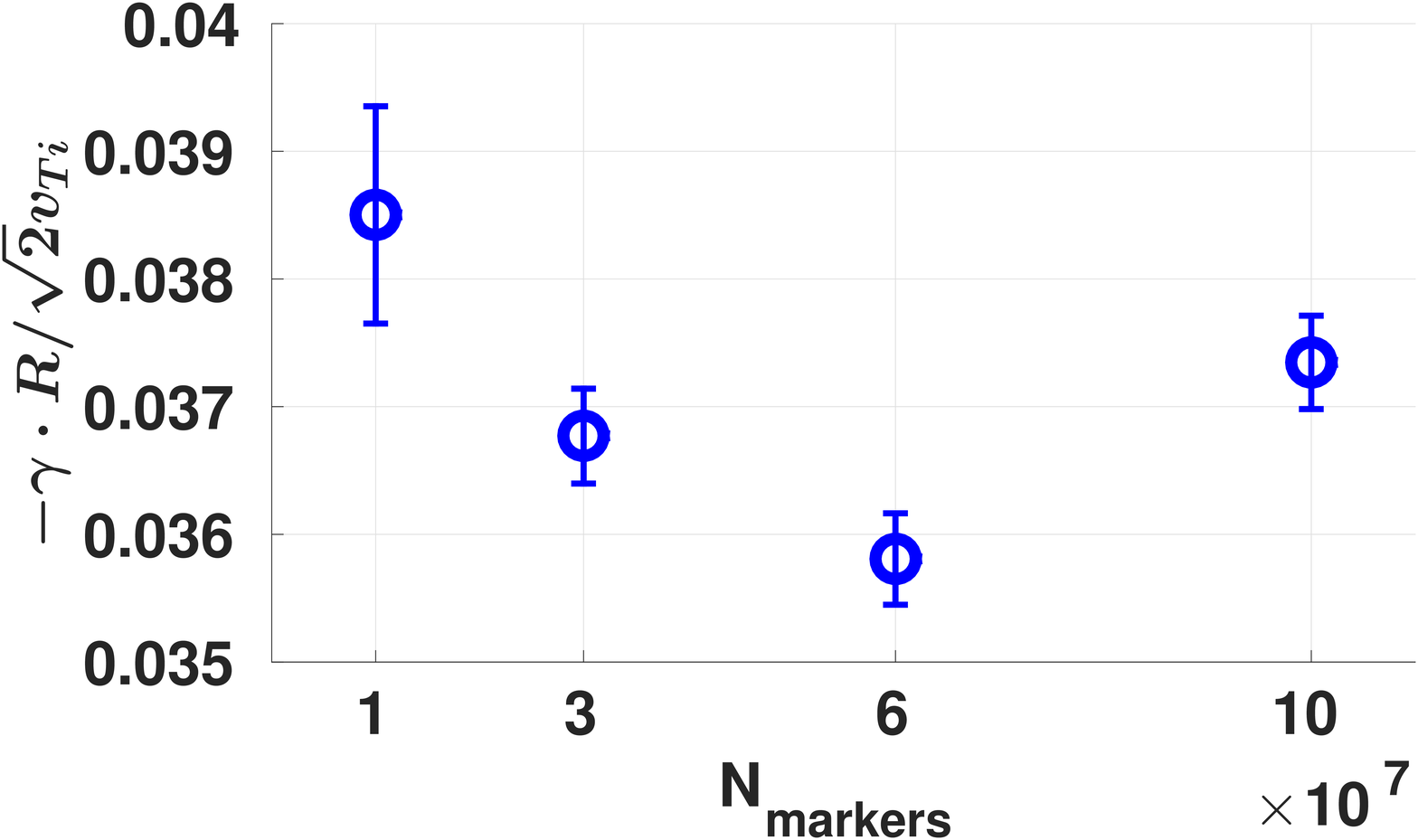}
\caption{\label{fig:ConvTests_nptot_g}}
\end{subfigure}
\caption{\label{fig:ConvTests} Convergence tests on the number of points in the radial space grid $n_s$ ($e = 1.60$, $q = 4.0$, $k_r \rho_i = 0.108$) (Fig. \ref{fig:ConvTests_ns_w} \ref{fig:ConvTests_ns_g}), on the normalized time step $dt_{norm}$ for different values of the plasma elongation $e = 1.00, 1.60$ ($q = 3.5$, $k_r \rho_i = 0.108$) (Fig. \ref{fig:ConvTests_dt_w} \ref{fig:ConvTests_dt_g}) and on the number of markers $N_{markers}$ ($e = 1.60$, $q = 4.0$, $k_r \rho_i = 0.108$) (Fig. \ref{fig:ConvTests_nptot_w} \ref{fig:ConvTests_nptot_g}).  }
\end{figure}

The transition from the simulations with the adiabatic electrons to the ones with the kinetic electrons applies additional restrictions on several numerical parameters such as the time step and the number of markers (see Fig. \ref{fig:ConvTests}).
In projects with adiabatic electrons the normalized time step $dt_{norm} = dt\cdot\omega_{ci}$ can be of the order of 20, but in case of the kinetic electrons it has to be significantly reduced till 2 because of the high parallel velocity of the passing electrons. Also electrostatic simulations of the kinetic electrons reveal high-frequency oscillations\cite{Lee87}. These oscillations can lead to numerical instabilities in case of low number of markers. To reduce their level we have passed to electromagnetic simulations with small values of electron beta that gave us an opportunity to keep the number of markers on the level of $10^7$.

The radial space step (or number of the points in the radial space grid) is determined by, among other parameters, the GAM wavenumber. To investigate the GAM dynamics with higher values of the radial wavenumber, we simulated a narrow poloidal ring near the edge instead of the full plasma cross-section to reduce the number of radial space points.

\section{Comparison with the code GENE}
\label{appendix:ComGENE}
\begin{figure}[t!]
\centering
\includegraphics[scale = 0.18]{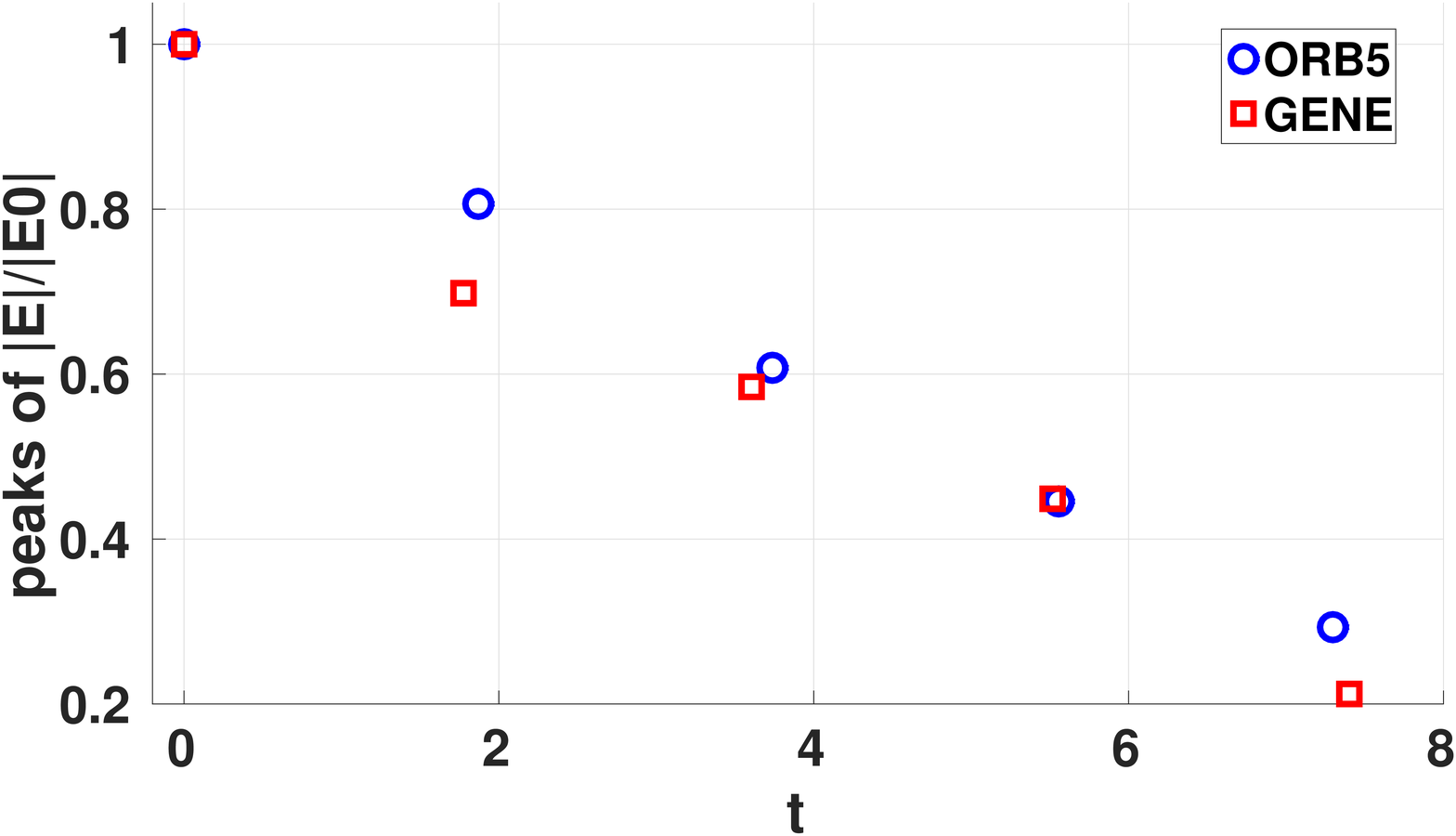}
\caption{\label{fig:GENE_peaks_kT10}Comparison between ORB5 and GENE for the case of the ion-electron temperature gradient $k_T = 10$. Here, the time is normalized to $R/\sqrt{2}v_{Ti}$.}
\end{figure}
A complete cross-code verification between the gyrokinetic ORB5 and GENE \cite{Jenko00, Gorler11} codes has already been done in Ref. \cite{Biancalani17} on the linear collisionless dynamics of the GAMs with adiabatic and kinetic electrons in the specific case of flat temperature profiles. 
For completeness, in this paper a comparison between these different codes is shown including the additional phase mixing physical effect, which is driven by non-flat temperature profiles.
The motivation behind this study is that although the linear physical models between ORB5 and GENE are equivalent \cite{Tronko17}, the numerical schemes are different. GENE is an Eulerian code, where the distribution function is not discretized with markers, but it is discretized on a 5D fixed grid in phase-space $(\yb{R}, v_{\parallel}, \mu)$, where $\yb{R}$ is the gyrocenter position, $v_{\parallel}$ is the parallel velocity, and $\mu$ is the magnetic momentum.

The simulation plasma parameters have been taken as in Sec. \ref{PhaseMixing} for both GENE and ORB5. A sinusoidal perturbation in the potential field is initialised, as defined in Sec. \ref{PARS_WOMIX} and is let evolved in time. In GENE, the radial box size is $60 \rho_s$. We have used 128 grid points in radial direction in order to have at least two points per ion Larmor radius. Along the field line 68 points have been used. In velocity space, 68 points and 128 equidistant symmetric grid points have been used for resolving respectively the $\mu$ and the $v_{\parallel}$ space. The velocity space domain has been fixed to 3 and 9 times the thermal velocity, respectively in the $v_{\parallel}$ and $\mu$ space. In order to avoid any recurrence problem, an hyperdiffusivity scheme has been used in the $v_{\parallel}$ direction. In Fig. \ref{fig:GENE_peaks_kT10} peaks of the flux-surface averaged radial electric field, measured at the radial position $s_0 = 0.90$ for the ion-electron temperature gradient $k_T = 10$, are shown for both GENE and ORB5. Half-decay time, calculated in ORB5, is $t_{1/2}^{orb}[R/\sqrt{2} v_{Ti}] = 4.9$ and for the case of GENE it is $t_{1/2}^{GENE}[R/\sqrt{2} v_{Ti}] = 4.8$.

\bibliographystyle{nar}
\bibliography{LinearGAM_bibl.bib}

\end{document}